\documentclass[twocolumn]{emulateapj}
\usepackage{graphicx}
\usepackage{dcolumn}

\newcommand       \be           {\begin{equation}}
\newcommand       \ee           {\end{equation}}

\newcommand       \K            {\,{\rm K}}
\newcommand       \cm           {\,{\rm cm}}
\newcommand       \s            {\,{\rm s}}

\newcommand       \erg          {\,{\rm erg}}

\newcommand       \pc           {\,{\rm pc}}

\newcommand       \gtsim        {\gtrsim}

\begin{document}

\title{Mid-Infrared Fine Structure Line Ratios in Active Galactic Nuclei Observed with Spitzer IRS: Evidence for Extinction by the Torus}

\author{R. P. Dudik\altaffilmark{1}, J. C. Weingartner\altaffilmark{1}, S. Satyapal \altaffilmark{1}, J. Fischer\altaffilmark{2}, C. C. Dudley \altaffilmark{2}, \& B. O'Halloran\altaffilmark{1}}

\altaffiltext{1}{George Mason University, Department of Physics \& Astronomy, MS 3F3, 4400 University Drive, Fairfax, VA 22030}

\altaffiltext{2}{Naval Research Laboratory, Remote Sensing Division, 4555 Overlook Ave SW, Washington DC, 20375}

\begin{abstract}
We present the first systematic investigation of the [NeV] (14$\mu$m/24$\mu$m) and [SIII] (18$\mu$m/33$\mu$m) infrared line flux ratios, traditionally used to estimate the density of the ionized gas, in a sample of 41 Type 1 and Type 2 active galactic nuclei (AGNs) observed with the Infrared Spectrograph on board {\it Spitzer}.  The majority of galaxies with both [NeV] lines detected have observed [NeV] line flux ratios consistent with or below the theoretical low density limit, based on calculations using currently available collision strengths and ignoring absorption and stimulated emission. We find that Type 2 AGNs have lower line flux ratios than Type 1 AGNs and that all of the galaxies with line flux ratios below the low density limit are Type 2 AGNs.  We argue that differential infrared extinction to the [NeV] emitting region due to dust in the obscuring torus is responsible for the ratios below the low density limit and we suggest that the ratio may be a tracer of the inclination angle of the torus to our line of sight. Because the temperature of the gas, the amount of extinction, and the effect of absorption and stimulated emission on the line ratios are all unknown, we are not able to determine the electron densities associated with the [NeV] line flux ratios for the objects in our sample.  We also find that the [SIII] emission from the galaxies in our sample is extended and originates primarily in star forming regions.  Since the emission from low-ionization species is extended, any analysis using line flux ratios from such species obtained from slits of different sizes is invalid for most nearby galaxies.

\end{abstract}


\section{Introduction}
		
	Mid-infrared (mid-IR) emission-line spectroscopy of active galactic nuclei (AGNs) is used to investigate the physical conditions of the dust-enshrouded gas that is in close proximity to the active nucleus.   In particular, many spectral lines are emitted in the so-called narrow-line region (NLR) of these objects which typically extends between tens to at most a thousand parsecs from the nucleus (Capetti, et al. 1995, 1997, Schmitt \& Kinney 1996, Falcke et al. 1998; Ferruit et al. 1999, Schmitt et al. 2003).  

The NLRs of AGNs have been studied extensively using optical spectroscopic observations.  However, there have been very few systematic studies of the NLR using infrared spectroscopic observations.  Infrared (IR) fine-structure emission lines have a number of special characteristics that have been regarded as distinct advantages, particularly in determining the electron density of the ionized gas very close to the central AGN.  Infrared spectroscopic observations allow access to fine-structure lines from ions with higher ionization potentials than the most widely used optical diagnostic lines. This is important in many AGNs, where a significant fraction of the line emission from lower ionization species can originate in gas ionized by star forming regions.  In addition, it is generally assumed that the density-sensitive infrared line ratios originate in gas with temperatures around 10$^4$ K and are less dependent on electron temperature variations, enabling a more straightforward determination of the electron density in the ionized gas.  Finally, it has long been assumed that the IR diagnostic line ratios are insensitive to reddening corrections--a serious impediment to optical and ultraviolet observations, particularly in the NLRs of AGNs, where the dust composition and spatial distribution are highly uncertain.  For these reasons, IR spectroscopic observations, especially since the era of the {\it Infrared Space Observatory} ({\it ISO}), have provided us with some of the most reliable tools for studying the NLRs in AGNs.  However, while there are clear advantages of mid-IR fine-structure diagnostics in studying the physical state of the ionized gas, very little work has been done to investigate their robustness in determining the gas densities of the NLRs in a large sample of AGNs.  The {\it Spitzer Space Telescope} Infrared Spectrometer (IRS), with its extraordinary sensitivity and spectral resolution, offers the opportunity to examine for the first time the physical state of NLR gas in a large sample of AGNs. 

	The focus of most previous comparative studies of the infrared fine-structure lines in AGNs has been on the excitation state of the ionized gas, in an effort to determine the existence and energetic importance of potentially buried AGNs and to constrain their ionizing radiation fields (Genzel et al. 1998, Lutz et al. 1999, Alexander \& Sternberg 1999, Sturm et al. 2002, Satyapal, Sambruna, \& Dudik 2004, Spinoglio et al. 2005). Remarkably, very little work has been done in the infrared on studying the line flux ratios traditionally used to probe the NLR gas densities in a significant number of AGNs.  We present in this paper the first systematic infrared spectroscopic study of the line flux ratios of [NeV] and [SIII] in order to 1) test the robustness of these line ratios as density diagnostics and 2) if possible, to probe the densities of the NLR gas in a large sample of AGNs.

\section{The Sample}

We searched the {\it Spitzer} archive for galaxies with an active nucleus and both high- and low-resolution Infrared Spectrometer (IRS; Houck et al. 2004) observations currently available.   Only those galaxies with indisputable optical, X-ray, or radio signatures of active nuclei (such as broad H$\alpha$ or X-ray or radio point sources) were included in our sample.  The sample includes three AGN subclasses: Seyferts, LINERs, and Quasars.  The galaxies in this sample span a wide range of distances (4 to 400 Mpc; median = 21 Mpc), Hubble types, bolometric luminosities (log (L$_{BOL}$) $\sim$ 40 to 46, median = 43), and Eddington Ratios (log(L/L$_{Edd}$) $\sim$ -6.5 to 0.3; median= -2.5).  The entire sample consists of 41 galaxies.  The basic properties of the sample are given in Table 1.  The black hole masses listed in Table 1 were derived using resolved stellar kinematics, if available, reverberation mapping, or by applying the correlation between optical bulge luminosity and central black hole mass determined in nearby galaxies only when the host galaxy was clearly resolved. Bolometric luminosities listed in Table 1 were calculated from the X-ray luminosities for most objects.  For Seyferts, the relationship L$_{BOL}$ = 10 $\times$ L$_X$ was adopted (Elvis 1994).  For LINERs\footnote[1]{We include all galaxies that are classified as LINERs using {\it either} the Heckman (1980) or Veilleux \& Osterbrock (1987) diagnostic diagrams.} we assumed L$_{BOL}$ = 34 $\times$ L$_X$, as derived from the spectral energy distribution of a sample of nearby LINERs from Ho (1999) (see also Dudik et al. 2005 and Satyapal et al. 2005).  The bolometric luminosities and black hole masses for quasars and radio galaxies were taken from Woo \& Urry (2002) and Marchesini, Celotti, \& Ferrarese (2004), respectively.  A detailed discussion of our methodology and justification of assumptions for determining black hole masses and bolometric luminosities for the various AGN classes represented in Table 1 can be found in Satyapal et al. (2005) and Dudik et al. (2005).  Table 1 also lists the AGN type (1 or 2) for the galaxies in our sample based on the presence or absence of broad (full width at half max (FWHM) exceeding 1000 km s$^{-1}$) Balmer emission lines in the optical spectrum.  We emphasize that the selection basis for the objects in our sample was on the availability of high resolution IRS {\it Spitzer} observations.  The sample should therefore not be viewed as complete in any sense.
\begin{table}
\fontsize{7pt}{9pt}\selectfont
\begin{center}
\begin{tabular}{lcccccc}
\multicolumn{7}{l}{{\bf Table 1: Properties of the Sample}}  \\ 
 \hline
\multicolumn{1}{c}{Galaxy} & \multicolumn{1}{c}{Distance} & Hubble & log & log & log & AGN \\

\multicolumn{1}{c}{Name} & \multicolumn{1}{c}{(Mpc)} & Type & 
 (M$_{\rm BH}$) & (L$_X$) & (L/L$_{Edd}$) & Type  \\

\multicolumn{1}{c}{(1)} & \multicolumn{1}{c}{(2)} & (3) & (4) & (5) & (6) & (7)\\ 
\hline
\multicolumn{7}{l}{{\it Seyferts}}\\
\hline
NGC4151 & 13 & SABab & 7.13$^a$ & 42.7$^b$ & -1.53 & 1$^r$\\
NGC1365 & 19 & SBb & 7.64$^b$ & 41.3$^d$ & -3.42 & 2$^s$\\
NGC1097 & 15 & SBb & $\cdots$ & 40.7$^e$ & $\cdots$ & $\cdots$\\
NGC7469 & 65 & SABa & 6.84$^a$ & 44.3$^a$ & 0.34 & 1$^t$\\
NGC4945 & 4 & SBcd & 7.35$^b$ & 42.5$^f$ & -1.97 & 2$^u$\\
Circinus & 4 & SAb & 7.72$^b$ & 42.1$^g$ & -2.74 & 2$^v$\\
Mrk 231 & 169 & SAc & 7.24$^c$ & 42.2$^h$ & -2.16 & $\cdots$\\
Mrk3 & 54 & S0 & 8.65$^a$ & 43.5$^a$ & -2.21 & 2$^w$\\
Cen A  & 3 & S0 & 7.24$^b$ & 41.8$^i$ & -2.54 & 2$^x$\\
Mrk463 & 201 & Merger & $\cdots$ & 43.0$^j$ & $\cdots$ & 2$^y$\\
NGC 4826 & 8 & SAab & 6.76$^b$ & $\cdots$ & $\cdots$ & $\cdots$\\
NGC 4725 & 16 & SABab & 7.40$^b$ & $\cdots$ & $\cdots$ & 2$^r$\\
1 ZW 1 & 245 & Sa & $\cdots$ & 43.9$^k$ & $\cdots$ & $\cdots$\\
NGC 5033 & 19 & SAc & 7.39$^b$ & 41.4$^l$ & -3.13 & 1$^r$\\
NGC1566 & 20 & SABbc  & 6.92$^a$ & 43.5$^a$ & -0.57 & 1$^t$\\
NGC 2841 & 9 & SAb & 8.21$^a$ & 42.7$^a$ & -2.64 & $\cdots$\\
NGC 7213 & 24 & SA0 & 7.99$^a$ & 43.30$^a$ & -1.79 & $\cdots$\\

\hline
\multicolumn{7}{l}{{\it LINERs}$^*$}\\
\hline

NGC4579 & 17 & SABb & 7.85$^b$ & 41.0$^b$ & -3.47 & $\cdots$\\
NGC3031 & 4 & SAab & 7.79$^b$ & 40.2$^b$ & -4.16 & $\cdots$\\
NGC6240 & 98 & Merger & 9.15$^b$ & 44.2$^b$ & -1.52 & 2$^z$\\
NGC5194 & 8 & SAbc & 6.90$^b$ & 41.0$^b$ & -2.43 & 2$^r$\\
MRK266NE & 112 & Merger & $\cdots$ & 40.9$^b$ & $\cdots$ & 2$^t$\\
NGC7552 & 21 & SBab & 6.99$^b$ & $\cdots$ & $\cdots$ & $\cdots$\\
NGC 4552 & 17 & $\cdots$ & 8.57$^b$ & 39.6$^b$ & -5.52 & $\cdots$\\
NGC 3079 & 15 & SBc & 7.58$^b$ & 40.1$^m$ & -4.05 & $\cdots$\\
NGC 1614 & 64 & SBc  & 6.94$^b$ & $\cdots$ & $\cdots$ & $\cdots$\\
NGC 3628 & 10 & SAb & 7.86$^b$ & 39.9$^n$ & -4.58 & $\cdots$\\
NGC 2623 & 74 & Pec & 6.83$^b$ & $\cdots$ & $\cdots$ & 2$^{aa}$\\
IRAS23128-5919 & 178 & Merger & $\cdots$ & 41.0$^b$ & $\cdots$ & 2$^{bb}$\\
MRK273 & 151 & Merger & 7.74$^b$ & 44.0$^o$ & -0.31 & 2$^t$\\
IRAS20551-4250 & 171 & Merger & 7.52$^c$ & 40.9$^b$ & -3.23 & $\cdots$\\
NGC3627 & 10 & SABb & 7.16$^b$ & 39.4$^p$ & -4.33 & 2$^r$\\
UGC05101 & 158 &  S & $\cdots$ & 40.9$^b$ & $\cdots$ & 1$^{cc}$\\
NGC4125 & 18 & E6 & 8.50$^b$ & 38.6$^b$ & -6.47 & $\cdots$\\
NGC 4594 & 10 & SAa & 9.04$^b$ & 40.1$^q$ & -5.47 & $\cdots$\\

\hline
\multicolumn{7}{l}{{\it Quasars}}\\
\hline

PG 1351+640 & 353 & $\cdots$ & 8.48$^a$ & 44.5$^a$ & -1.08 & $\cdots$\\
PG 1211+143 & 324 & $\cdots$ & 7.49$^a$ & 44.8$^a$ & 0.22 & 1$^t$\\
PG 1119+120 & 201 & $\cdots$ & $\cdots$ & $\cdots$ & $\cdots$ & 1$^y$\\
PG 2130+099 & 252 & Sa & 7.74$^a$ & 44.47$^a$ & -0.37 & 1$^y$\\
PG 0804+761 & 400 & $\cdots$ & 8.24$^a$ & 44.93$^a$ & -0.41 & 1$^{dd}$\\
PG 1501+106 & 146 & E  & $\cdots$ & $\cdots$ & $\cdots$ & 1$^y$\\

\hline
\end{tabular}
\end{center}
{\scriptsize{\bf Columns Explanation:} Col(1):Common Source Names; Col(2):  Distance (for H$_0$= 75 km s$^{-1}$Mpc$^{-1}$); Col(3): Morphological Class; Col(4): Mass of central black hole in solar masses; Col(5): Log of the hard  X-ray luminosity (2-10keV) in erg s$^{-1}$. Col(6): log of the Eddington Ratio. (* = We include all galaxies that are classified as LINERs using {\it either} the Heckman (1980) or Veilleux \& Osterbrock (1987) diagnostic diagrams. Col(6): AGN type based on the presence or absence of broad Balmer emission lines.)}
{\scriptsize{\bf References:}$^a$Woo \& Urry 2002, $^b$ Satyapal et al. 2005, $^c$ Tacconi et al. 2002, $^d$ Risaliti et al. 2005, $^e$ Terashima et al. 2002, $^f$ Guainazzi et al. 2000, $^g$ Smith \& Wilson 2001, $^h$ Gallagher et al. 2002, $^i$ Evans et al. 2004, $^j$ Imanishi \& Terashima et al. 2004, $^k$Gallo et al. 2004 , $^l$ Terashima et al. 1999, $^m$ Cappi et al. 2006, $^n$ Roberts, Schurch, \& Warwick 2001, $^o$ Balestra et al. 2005, $^p$Georgantopoulos et al 2002, $^q$ Dudik et al. 2005, $^r$ Ho et al. 1997, $^s$ Storchi-Bergmann, Mulchaey, \& Wilson, 1992, $^t$ Veron-Cetty \& Veron 2003, $^u$ Marconi et al. 2000, $^v$Oliva et al. 1994, $^w$ Khachikian  \& Weedman 1974, $^x$ Veron-Cetty \& Veron 1986$^y$ Dahari \& De Robertis 1988, $^z$ Andreasian, Khachikian, \& Ye, 1987, $^{aa}$ Laine et al. 2003, $^{bb}$ Duc, Mirabel, \& Maza 1997, $^{cc}$ Sanders et al. 1988, $^{dd}$ Thompson 1992. }
\end{table}


\section{Data Analysis and Results}
	We extracted archival spectral data obtained using the short-wavelength, low-resolution module (SL2, 3.6''$\times$57'', $\lambda$ = 5.2-7.7$\mu$m) and both the short-wavelength, high-resolution (SH, 4.7''$\times$11.3'', $\lambda$  =  9.9-19.6$\mu$m) and long-wavelength, high-resolution (LH, 11.1''$\times$22.3'', $\lambda$ = 18.7-37.2$\mu$m) modules of IRS.  

	The data presented here were preprocessed by the IRS pipeline (version 13.0) at the {\it Spitzer} Science Center (SSC) prior to download.  Preprocessing includes ramp fitting, dark-sky subtraction, droop correction, linearity correction, flat-fielding, and flux calibration\footnote[2]{See {\it Spitzer} Observers Manual, Chapter 7, http://ssc.spitzer.caltech.edu/documents/som/irs60.pdf}.  The {\it Spitzer} data were further processed using the SMART v. 5.5.7 analysis package (Higdon et al. 2004).  The slit for the SH and LH modules is too small for background subtraction to take place and separate SH or LH background observations do not exist for any of the galaxies in this sample.  For the SL2 module, background subtraction was done using either a designated background file when available or the interactive source extraction option.   In the case of the latter, the exact position of the slit on the host galaxy was first checked using {\it Leopard}, the data archive access tool available from the SSC.  The source was then carefully defined according to the boundary of the slit and the edge of the host galaxy.  The background was defined at the edge of the slit, where no other obvious source was present.  In some cases, the slit was enveloped in the host galaxy and background subtraction could not take place.  For both high and low resolution spectra, the ends of each order were manually cut from the rest of the spectrum.  

	The 41 observations presented in this work are archived from various programs, including the {\it SINGS} Legacy Program, and therefore contain both mapping and staring observations.   All of the staring observations were centered on the nucleus of the galaxy.  The SH, LH, SL2 staring observations include data from two slit positions overlapping by one third of a slit.   In order to isolate the nuclear region in the mapping observations so that we might compare them to the staring observations, we extracted only those 3 overlapping slit positions coinciding with either radio or {\it 2MASS} nuclear coordinates.  Because the slits in both the mapping and staring observations occupy distinctly different regions of the sky, the slits cannot be averaged unless the emission originates from a compact source that is contained entirely in each slit.  Therefore the procedure for flux extraction was the following: 1) If the fluxes measured from the two slits differed by no more than the calibration error of the instrument, then the fluxes were averaged; otherwise, the slit with the highest measured line flux was chosen.  2) If an emission line was detected in one slit, but not in the other, then the detection was selected.  This is true for all of the high and low resolution staring and mapping observations.

In Tables 2 and 3 we list the line fluxes and statistical errors from the SH and LH observations for the [NeV] 14.3$\mu$m and 24.3$\mu$m lines, the [SIII] 18.1$\mu$m and 33.5$\mu$m lines, as well as the 6.2$\mu$m PAH emission feature. For all galaxies with previously published fluxes, we list in Tables 2 and 3 the published flux values. Our values differ by no more than a factor of 1.9, much less in most cases, from the Weedman et al. (2005) or Armus et al. (2004, 2006) published values.  These differences can be attributed to differences in the pipeline used for preprocessing.  In all cases detections were defined when the line flux was at least 3$\sigma$.  For the absolute photometric flux uncertainty we conservatively adopt 15\%, based on the assessed values given by the {\it Spitzer} Science Center (SSC) over the lifetime of the mission.\footnote[3]{See {\it Spitzer} Observers Manual, Chapter 7, (http://ssc.spitzer.caltech.edu/documents/som/som7.1.irs.pdf and IRS Data Handbook (http://ssc.spitzer.caltech.edu/irs/dh/dh20$_{}$v2.pdf, Chapter 7.2 } This error is calculated from multiple observations of various standard stars throughout the {\it Spitzer} mission by the SSC.  The dominant component of the total error arises from the uncertainty at mid-IR wavelengths in the stellar models used in calibration and is systematic rather than Gaussian in nature.  We note that the spectral resolution of the SH and LH modules of IRS ($\lambda$ / $\Delta$$\lambda$ $\sim$ 600) is insufficient to resolve the velocity structure for most of the lines.  There are a few galaxies which do show slightly broadened [NeV] line profiles (FWHM $\sim$ 200 - 1200 km s$^{-1}$).  These results will be discussed in a future paper.

	Abundance-independent density estimates can readily be obtained using infrared fine-structure transitions from like ions in the same ionization state with different critical densities.  The density diagnostics available in the IRS spectra of our objects are: [NeV] 14.32$\mu$m, 24.32 $\mu$m (n$_{crit}$ $\sim$ 4.9 $\times$ 10$^4$ cm$^{-3}$, and 2.7 $\times$ 10$^4$  cm$^{-3}$, where n$_{crit}$ = A$_{ul}$/$\gamma$$_{ul}$, with A$_{ul}$ the Einstein A coefficient and $\gamma$$_{ul}$ the rate coefficient for collisional de-excitation from the upper to the lower level), [NeIII] 15.55$\mu$m, 36.04 $\mu$m (n$_{crit}$ $\sim$ 3 $\times$ 10$^5$ cm$^{-3}$, and 5 $\times$ 10$^4$ cm$^{-3}$, Giveon et al. 2002), and  [SIII]18.71$\mu$m, 33.48 $\mu$m (n$_{crit}$ $\sim$ 1.5 $\times$ 10$^4$ cm$^{-3}$, and 4.1 $\times$ 10$^3$ cm$^{-3}$).   The results are very insensitive to the shape of the ionizing continuum.  Since the [NeIII] 36$\mu$m line was either not detected or was outside the wavelength range of the LH module in virtually all galaxies, we omit any analysis of the [NeIII] line ratio from this work.

\section{The [NeV] Line Flux Ratios}

In Figure 1 we plot the calculated $14\micron/24\micron$ line luminosity ratio as a function of electron density $n_e$ for gas temperatures $T = 10^4 \K$, $10^5 \K$, and 10$^6 \K$.  We include only the five levels of the ground 2s$^2$2p$^2$ configuration and neglect absorption and stimulated emission.  The results are nearly identical if only the lowest three levels of the ground term are included.  We adopt collision strengths from Griffin \& Badnell (2000) and radiative transition probabilities from Galavis, Mendoza, \& Zeippen (1997).  
\begin{figure}[htbp]
{\includegraphics[width=8cm]{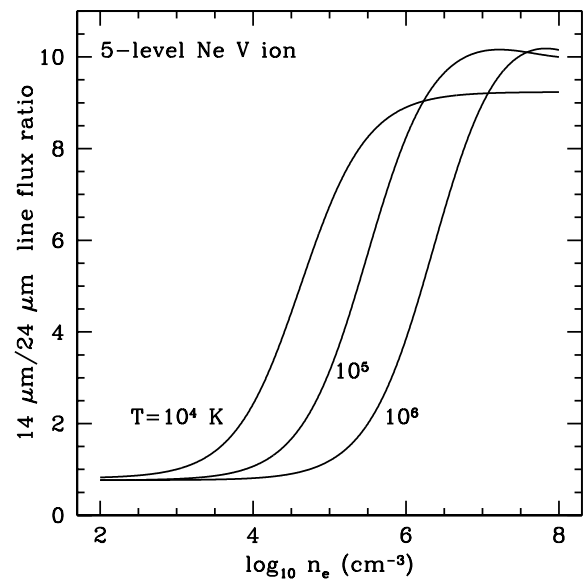}}\\
\caption[]{[NeV] $14\micron/24\micron$ line flux ratio versus electron density, $n_e$, for gas temperatures $T=10^4 \K$, $10^5 \K$, $10^6 \K$}
\end{figure}

In Table 2, we list the observed [NeV] line flux ratios and their associated calibration uncertainties.  In calculating the upper and lower limits on the ratios, $R_{MAX}$ and $R_{MIN}$, shown in Table 2, we did not propagate the errors in quadrature as would be appropriate for statistical uncertainties, but propagated them as follows:
\begin{eqnarray} 
R_{MAX} & = & \frac{F{\rm [NeV]}_{14} + 0.15 (F{\rm [NeV]}_{14})}{F{\rm [NeV]}_{24} - 0.15 (F{\rm [NeV]}_{24})} \\
&&\nonumber\\
R_{MIN} & = & \frac{F{\rm [NeV]}_{14} - 0.15 (F{\rm [NeV]}_{14})}{F{\rm [NeV]}_{24} + 0.15 (F{\rm [NeV]}_{24})} 
\end{eqnarray}
We note that this is conservative, since some components of the calibration errors should cancel in the ratio.  Both line fluxes were measured for 19 galaxies.  In what follows we compare the line flux ratios measured in all but one, MKN 266, for reasons that are discussed in detail in Section 5.2.  Of these 18 AGNs, 13 have ratios that are consistent with the low density limit to within the uncertainties, while only 2, both Type 1, have ratios significantly above it.  The remaining 3, all Type 2, have ratios significantly below the low-density limit.  Interestingly, we note that a similar range of ratios was also measured with the ISO SWS (Sturm et al. 2002, Alexander et al. 1999).  There are several possible explanations for this finding.  The observed, unphysically low ratios could result from artifacts introduced by variations in the slit sizes from which the line fluxes are obtained, from calibration uncertainties, or from substantial mid-IR extinction.  Alternatively, perhaps important physical processes were neglected in calculating the theoretical ratios.  In addition, errors in the collisional rate coefficients for the [NeV] transitions associated with the mid-infrared lines may be important.  We explore these scenarios in the following sections.

{\bf Observational Effects:}  Because the IRS LH slit is larger than the SH slit, if the [NeV] emission is extended, or multiple AGNs are present, the 14/24 $\mu$m line ratio will be artificially reduced.  However, since the ionization potential of [NeV] is $\sim$ 97 eV, we expect that the [NeV]-emitting gas is ionized by the AGN radiation field only and is concentrated very close to the central source.  Virtually all of the [NeV] fluxes presented in this work were obtained from IRS staring observations.  Thus it is impossible to determine whether the emission is extended using {\it Spitzer} observations alone. However, a number of galaxies have been observed at 14 and 24 $\mu$m by ISO.  In Table 2 we list in addition to our {\it Spitzer} [NeV] fluxes, all available [NeV] fluxes from ISO.  The ISO aperture at 14 and 24 $\mu$m (14''$\times$27'') is much larger than either the SH or LH slits.  In Figure 2 we plot the ratio of the [NeV] flux measured by ISO to that measured by {\it Spitzer} for both the 14 and 24 $\mu$m lines.  The ranges of the [NeV] line flux ratios are consistent with the instrument uncertainties and are similar for all galaxies in the sample.  Only the 14$\mu$m ratio for Mrk 266 falls outside of the expected range.  This strongly suggests that the [NeV] emission is indeed compact and originates in the NLR and that the ratios are {\it not} affected by aperture variations, except for Mrk 266 which is discussed in detail in Section 5.2.   

\begin{figure}[htbp]
{\includegraphics[width=9cm]{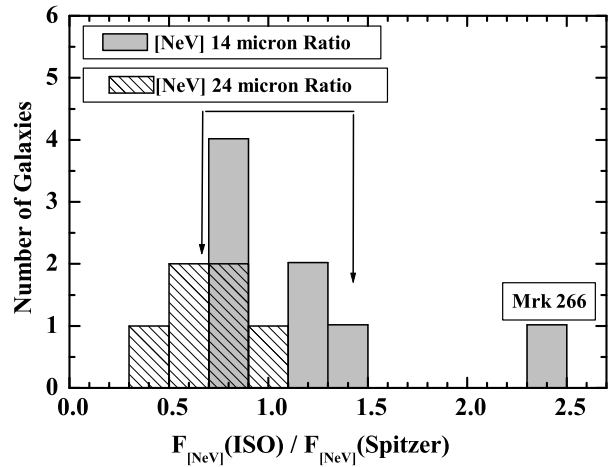}}\\
\caption[]{Ratio of the {\it ISO} to {\it Spitzer} [NeV] 14$\mu$m and 24$\mu$m fluxes for those galaxies with overlapping observations.  The range indicated with arrows is that corresponding to the absolute flux calibration for {\it ISO} (20\%) and {\it Spitzer} (15\%).  Within the calibration uncertainties of the instrument, the [NeV] fluxes are virtually the same for all of the galaxies except Mrk 266 (See Section 6.2).  This strongly suggests that the [NeV] emission is compact and originates in the NLR.  We note that Sturm et al. 2002 find that the [NeV] 24$\mu$m detection for NGC 7469 is questionable.  The {\it ISO} to {\it Spitzer} ratio for this galaxy (0.43) is the lowest shown here.}
\end{figure}

	If the data were affected by aperture variations we would expect to see an overall systematic increase of the 14$\mu$m/24$\mu$m line ratio with distance(See Figure 3). The Spearman rank correlation coefficient (r$_S$, Kendell \& Stuart 1976) corresponding to this plot is -0.069 (with a probability of chance correlation of 0.78), where a coefficient of 1 or -1 indicates a strong correlation and a coefficient of 0 indicates no correlation. Thus we find that there is no correlation between the [NeV] ratio and distance in our sample.  However this does not completely rule out aperture effects, if the size of the [NeV] emitting region increases with the bolometric luminosity of the AGN and the sample displays a significant trend in bolometric luminosity with distance.  In this case, a correlation between the [NeV] ratio and distance would not be apparent since aperture variations would affect all galaxies in the same way, regardless of distance.  However this scenario is unlikely since the size of the [NeV]-emitting region would have to increase proportionately with distance in order to remain extended beyond the slit for all galaxies.  Nevertheless, we checked for this possibility, both by examining the [NeV] ratio vs. bolometric luminosity and by plotting the ratio vs. distance, binning the galaxies according to their bolometric luminosity.  We find neither to be correlated over 5 orders of magnitude in L$_{BOL}$.  Thus, in the case of the [NeV] line flux ratio, we find no indication that ratios below the low density limit are artifacts of aperture effects.

\begin{figure}[htbp]
{\includegraphics[width=9cm]{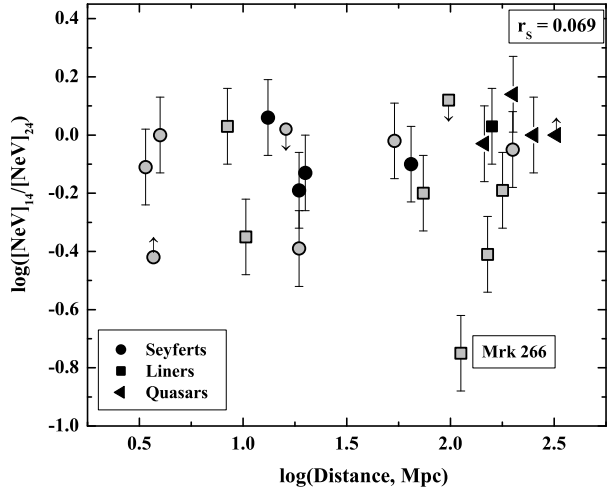}}\\
\caption[]{The [NeV] 14$\mu$m/24$\mu$m ratio as a function of distance.  Open symbols signify Type 1 AGNs, Filled symbols signify Type 2 AGNs.  The error bars shown here mark the calibration uncertainties on the line ratio.  If the ratio were indeed affected by aperture variations we would expect a systematic increase of the ratio with distance.  As can be seen here, this is not the case, and we find no indication that the low ratio is attribuable to aperture effects.}
\end{figure}

We point out that the [NeV] 24$\mu$m IRS line fluxes in the small overlapping sample plotted in Figure 2 are systematically higher than the corresponding {\it ISO}-SWS fluxes, despite the smaller IRS slit.  This indicates that one or both of the instruments is affected by systematic errors more severe than are indicated by the calibration uncertainty estimates.  The SWS band 3D that includes the [NeV] 24$\mu$m line was characterized by strong fringing effects that when combined with the narrow range of the line scan mode introduced sometimes large uncertainties in the baseline fitting, and therefore the line flux measurement accuracy.  In contrast, the baseline fitting over the entire {\it Spitzer} IRS SH and LH full spectra can be much more accurate.  Moreover pointing accuracy and stability are an order of magnitude improved over that obtained by ISO.  We therefore assume in the sections that follow that the adopted conservative {\it Spitzer} IRS calibration uncertainties are accurate characterizations of the IRS measurements.  Importantly, {\it regardless of which instrument is used, [NeV] ratios consistent with the low density limit have been observed in a number of sources with both {\it ISO}} (e.g. Sturm et al. 2002 NGC 1365, NGC 7582, NGC4151, NGC 5506; Alexander et al. 1999, NGC 4151) and {\it Spitzer} (Weedman et al. 2005, Haas et al. 2005, and this work).

{\bf Extinction:}   We consider the possibility that mid-IR differential extinction toward the [NeV]-emitting regions is responsible for the low [NeV] line ratios.  Adopting the low-density limit (LDL) for the intrinsic value of the ratio ([NeV]14$\mu$m/24$\mu$m $\sim$0.83 for n$_e$$\leq$200 cm$^{-3}$) for galaxies with ratios below the LDL, the observed line ratio gives a {\it lower limit} to the extinction, for a given MIR extinction curve.  We examined the visual extinctions corresponding to the mid-IR differential extinction derived using three separate extinction curves: 1) the Draine (1989) extinction curve amended by the more recent ISO SWS extinction curve toward the Galactic center for 2.5-10$\mu$m (Lutz et al. 1996), 2) the Chiar \& Tielens (2006) extinction curve for the Galactic Center using 2.38-40$\mu$m ISO SWS observations of a bright IR source in the Quintuplet cluster (GCS3-I) 3) the Chiar \& Tielens (2006) extinction curve for the local ISM using 2.38-40$\mu$m ISO SWS observations of a WC-type Wolf-Rayet (WR) star (WR98a). The Draine (1989) and Lutz et al. (1996) extinction curve yields A$_{V}$ $\sim$ 3 to 99 mag (See Table 2).  However, these values result from an extinction law that is unexplored beyond 10$\mu$m.  The Chiar \& Tielens (2006) Galactic center extinction curve cannot explain the observed [NeV] ratios since the extinction at 24$\mu$m is greater than the extinction at 14$\mu$m, so we do not discuss it further.  The visual extinction resulting from their local ISM extinction curve is unrealistically high (A$_V$$_{median}$=500mag).  The calculated extinction obtained using the Draine (1989), Lutz et al. (1996), and Chiar \& Tielens (2006) local ISM extinction curves are given in Table 2.

The A$_V$ derived from the two extinction curves described above are extremely high in many cases. Even if the extinction is calculated from the upper limit on the ratio to the LDL for the three galaxies whose upper limits are below the LDL, the corresponding visual extinction is still very high (for the Draine 1989 and Lutz 1996 extinction curve A$_V$ = 21, 26, and 30 mag for these three galaxies; for the Chiar and Tielens extinction curve A$_V$ = 260, 330, and 370 mag).  However, we caution the reader that the actual {\it value} for extinction is highly uncertain.  Indeed very little is known about the 8-40$\mu$m extinction curve in AGNs.  Specifically, the 10 and 18 $\mu$m silicate features in this band are the source of inconsistency.  Even within the AGN class, extinction may vary dramatically from 8-40$\mu$m because of variations in the silicate features due to differences in grain size, porosity, shape, composition, abundance, and location in each galaxy.  Hao et al. (2005) show that in five AGNs (4 of which are in our sample), both silicate features vary considerably in strength and width.  Sturm et al. (2005) also show that the standard ISM silicate models do not accurately fit NGC 3998, a LINER with silicate emission. Sturm et al. (2005) suggest that increased grain size and possibly the presence of crystalline silicates such as clino-pyroxenes may improve the fit, but that clearly circumnuclear dust in AGNs has very different properties than dust in the Galactic ISM (see also Maiolino et al. 2001a, 2001b, but Weingartner \& Murray 2002 for an alternative view).  Chiar \& Tielens (2006) even show that the GC observations and the local ISM observations within the Galaxy deviate from each other most dramatically in the wavelength region between the two silicate absorption features.  In their observations, this is the region between $\sim$ 12-15$\mu$m -directly overlapping with the 14$\mu$m values in which we are interested.  Because of irregularity of the silicate features in the mid-IR, it is very difficult to interpret the true extinction there.  Moreover, in addition to the uncertainty in the extinction law, the geometry of the obscuring material is unknown and can vary substantially from galaxy to galaxy.  The most that can be said here for the galaxies with ratios below the LDL is that if extinction is responsible for the low ratios, then the extinction must be less at 24$\mu$m than at 14$\mu$m.  

{\bf Physical Processes:} It is possible that important physical processes have been neglected in calculating the [NeV] line luminosity ratio as a function of electron density shown in Figure 1.  We consider three physical processes that may affect the line ratios:  

(1)  A source of gas heating in addition to photoionization 
(e.g., shocks, turbulence) that may yield gas temperatures substantially 
higher than $10^4 \K$.  As can be seen in Figure 1, higher gas temperatures do not yield significantly lower line ratios in the low-density limit, but could explain the generally low values of the ratios that lie above the LDL.

(2)  Pumping from the ground term to the first excited term, e.g., by 
O III resonance lines.  The specific energy density required for this
to significantly affect the line ratio exceeds $10^{-14} \erg \cm^{-3} 
\, {\rm Hz}^{-1}$, which is implausibly large by orders of magnitude.

(3)  Absorption and stimulated emission within the ground term, which could be important if, e.g., a large quantity of warm dust yielding copious 24$\mu$m continuum emission is located close to the [NeV]-emitting region.  Figures 4a through
4e show the line ratio as a function of the specific energy
density at $24 \micron$, $u_{\nu}(24\micron)$.  We display results for
electron density $n_e = 10^2$, $10^3$, $10^4$, $10^5$, and $10^6 \cm^{-3}$;
gas temperature $T = 10^4$, $10^5$, and $10^6 \K$; and ratio of the specific
energy density at 14 and $24 \micron$, 
$u_{\nu}(14 \micron)/u_{\nu}(24 \micron) = 0.4$, 1.0, and 1.8 (values were chosen to reproduce the observed range of the $14\micron/24\micron$ continuum flux ratios; see Section 5). 

For the moment, assume that the NeV is located sufficiently far from the 
source of the 14 and 24$\mu$m continuum emission to treat the source
as a point.  If hot dust within or near the inner edge of the torus is 
responsible for this emission, then this assumption
requires that the distance to the 
NeV, $r_{\rm Ne}$, be large compared with the dust sublimation radius, 
$r_{\rm sub} \sim 1 \pc \, L_{\rm bol, \, 46}^{1/2}$ (Ferland et al.~2002);
$L_{\rm bol, \, 46}$ is the bolometric luminosity in units of $10^{46} \erg
\s^{-1}$.  In this case, 
we can obtain a simple estimate of $u_{\nu}(24 \micron)$ at the location of
the [NeV]-emitting region from the observed specific flux 
$F_{\nu}(24 \micron)$, the distance to the galaxy $D$, and $r_{\rm Ne}$:
\be
u_{\nu}(24 \micron) \sim \frac{1}{c} F_{\nu}(24 \micron) \left( \frac{D}
{r_{\rm Ne}} \right)^2~~~.
\ee
With $r_{\rm Ne} = 100 \pc$, $u_{\nu}(24 \micron)$ estimated in this way ranges
from $\approx 10^{-24} \erg \cm^{-3} \, {\rm Hz}^{-1}$ to somewhat less
than $10^{-20} \erg \cm^{-3} \, {\rm Hz}^{-1}$ for the galaxies in our sample.
These can be compared 
to the results of H\"{o}nig et al.~(2006), who modeled the infrared emission 
from clumpy tori.  They presented plots of $F_{\nu}$ at a distance of 
10$\,$Mpc from an AGN with bolometric luminosity $L_{\rm bol} = 4 \times 
10^{45} \erg \s^{-1}$.  Extrapolating to a distance of $100 \pc$, we find
$u_{\nu}(24 \micron)$ as large as a few times $10^{-21} \erg \cm^{-3} \, 
{\rm Hz}^{-1}$, close to the estimate for the most luminous AGN in our 
sample.  

From Figures 4a through 4e, we see
that the infrared continuum can only reduce the line ratio significantly 
at $r_{\rm Ne} \approx 100 \pc$ if $T \gtsim 10^5 \K$ when 
$n_e \sim 10^2 \cm^{-3}$ and $T \gtsim 10^6 \K$ when
$n_e \sim 10^3 \cm^{-3}$.
However, the NeV, as a high-ionization species, may lie closer to the 
central source than does the bulk of the narrow line region.  If 
$r_{\rm Ne} \approx 10 \pc$, then $u_{\nu}(24 \micron)$ increases by a factor
$\sim 100$.  
In this case, the observed low line ratios can be explained by this
mechanism with $T \sim 10^4 \K$, if $n_e \sim 10^2 \cm^{-3}$.  Higher values 
of electron density would require higher gas temperatures.  
										\begin{figure*}[htbp]
\begin{center}
{\includegraphics[width=7cm]{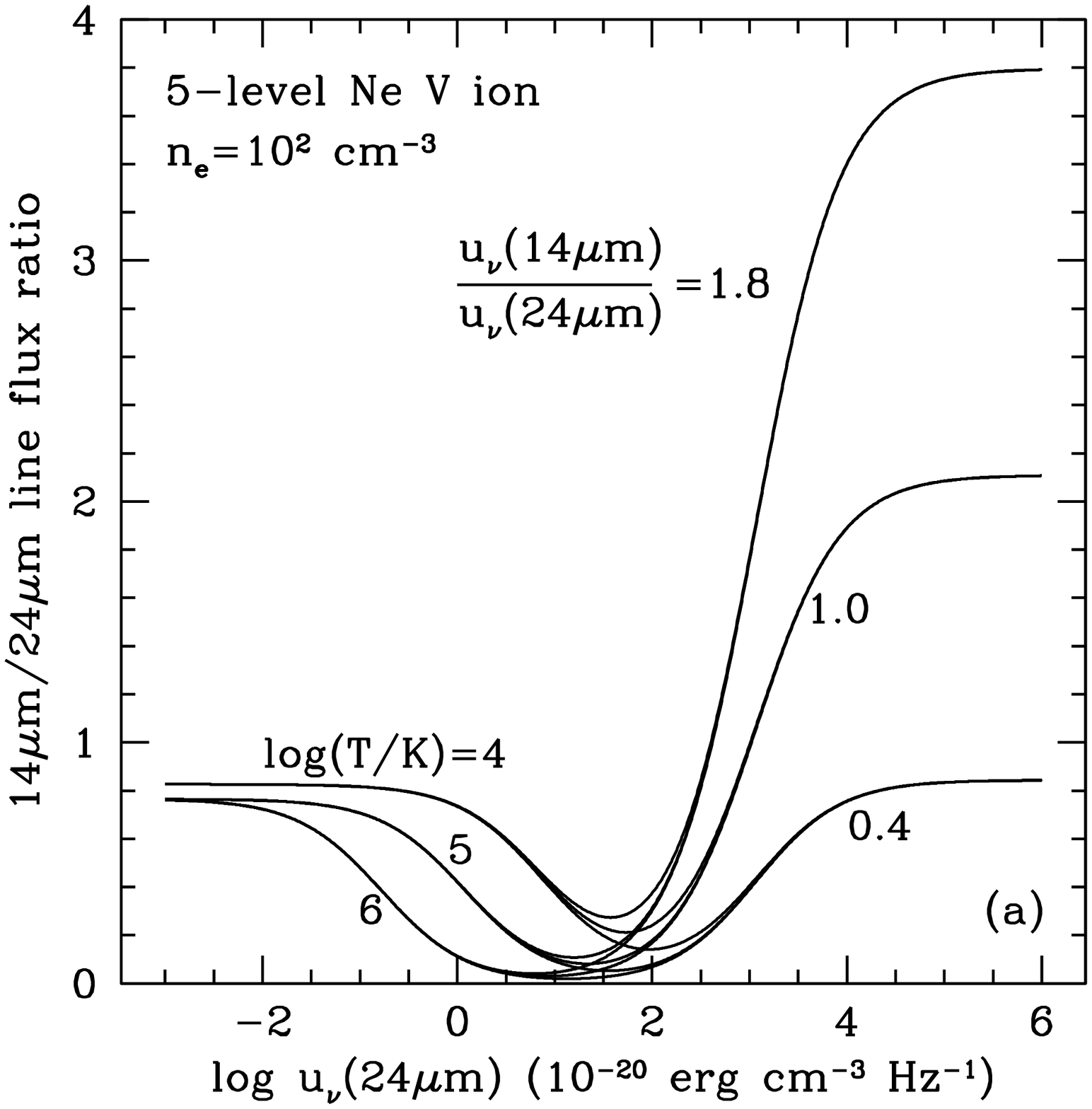}}{\includegraphics[width=7cm]{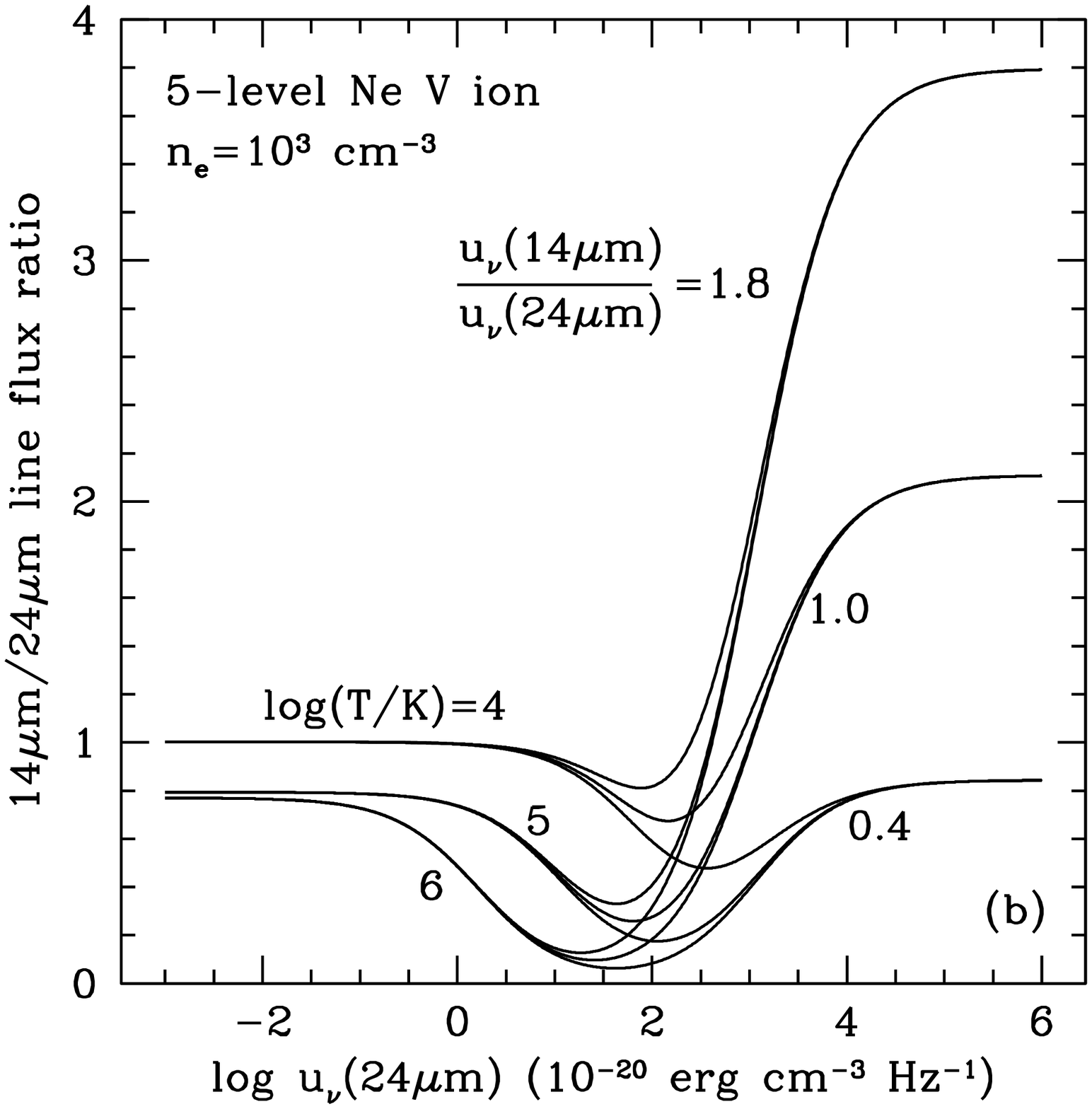}}\\{\includegraphics[width=7cm]{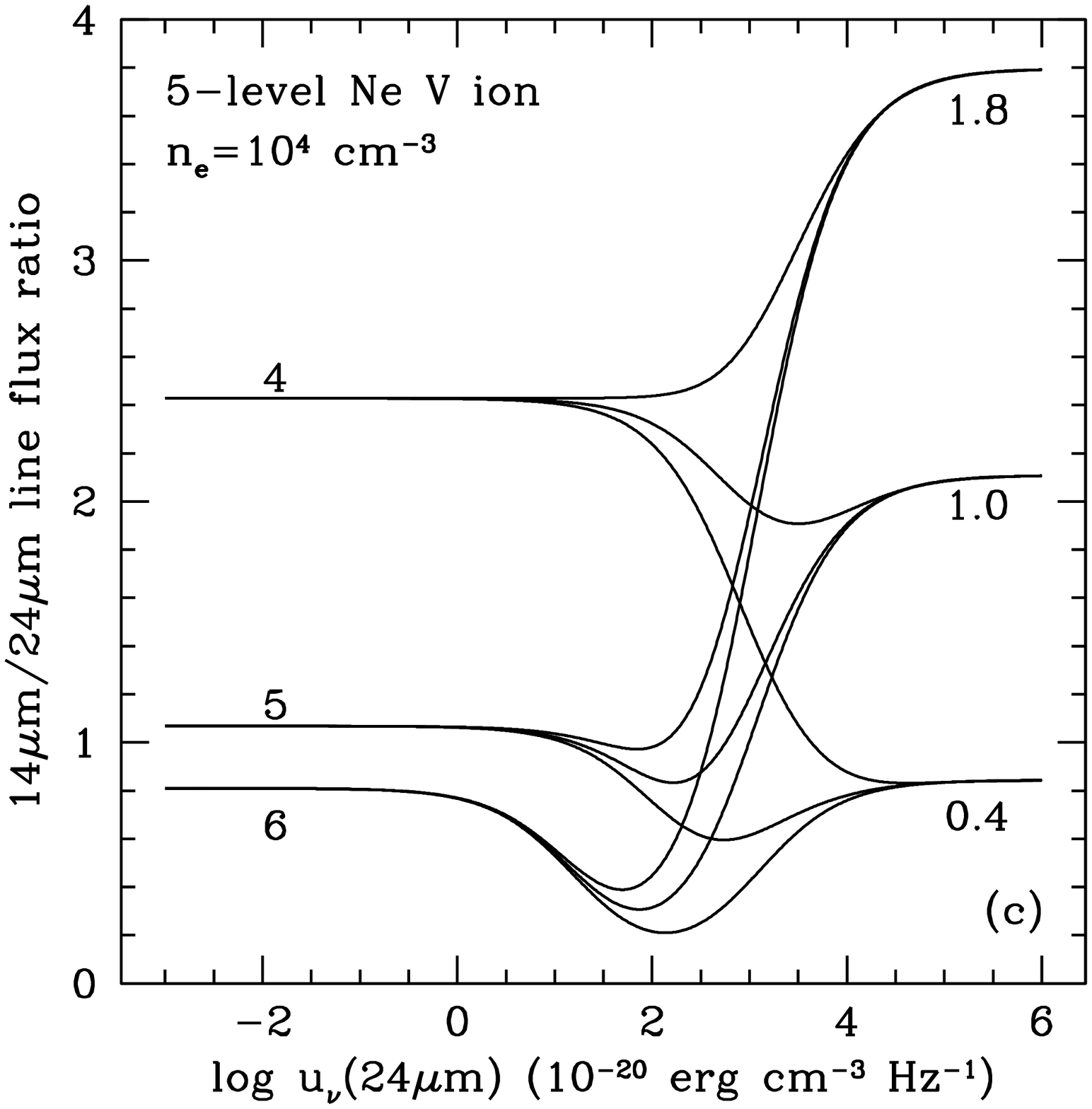}}{\includegraphics[width=7cm]{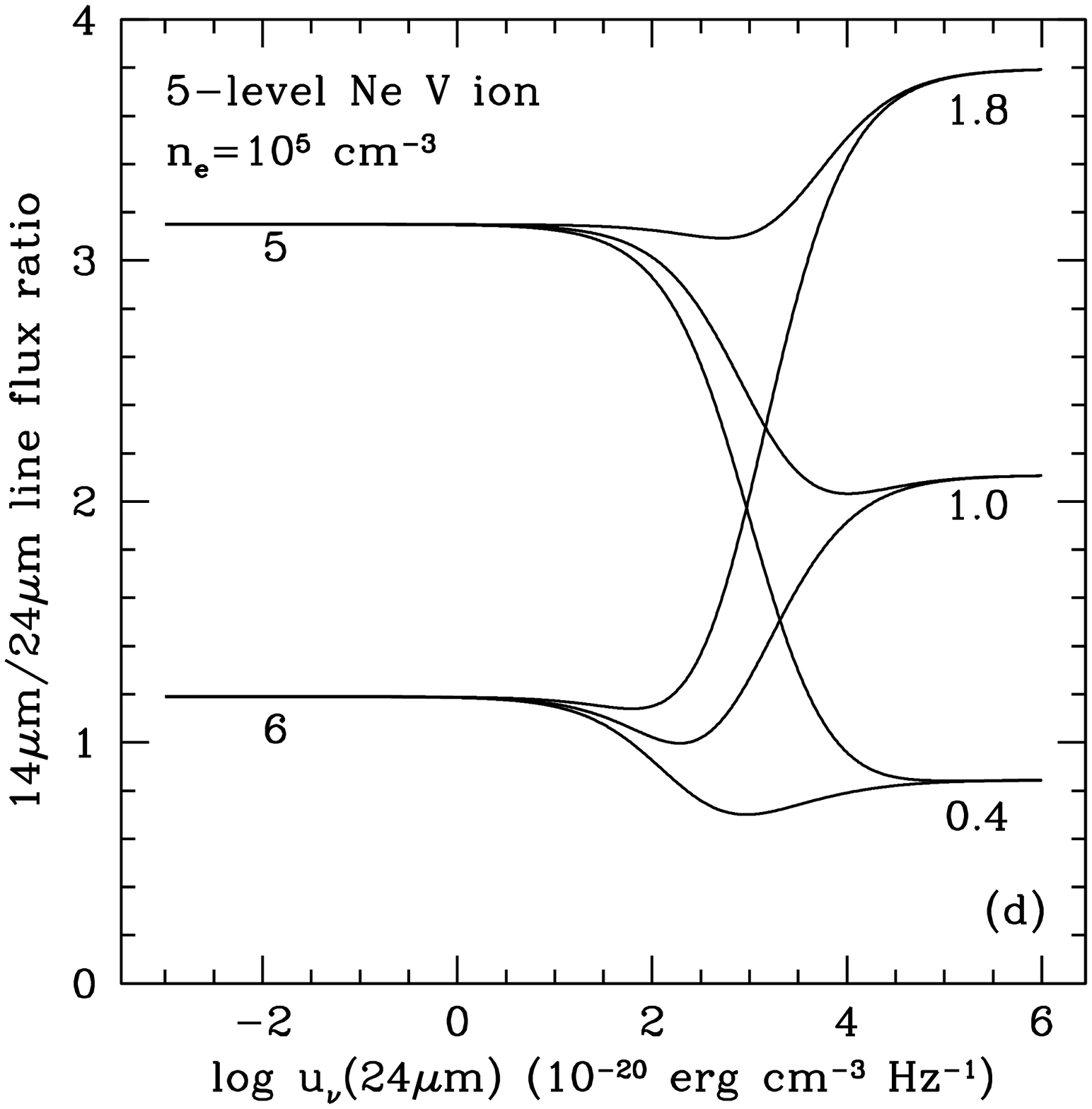}}\\{\includegraphics[width=7cm]{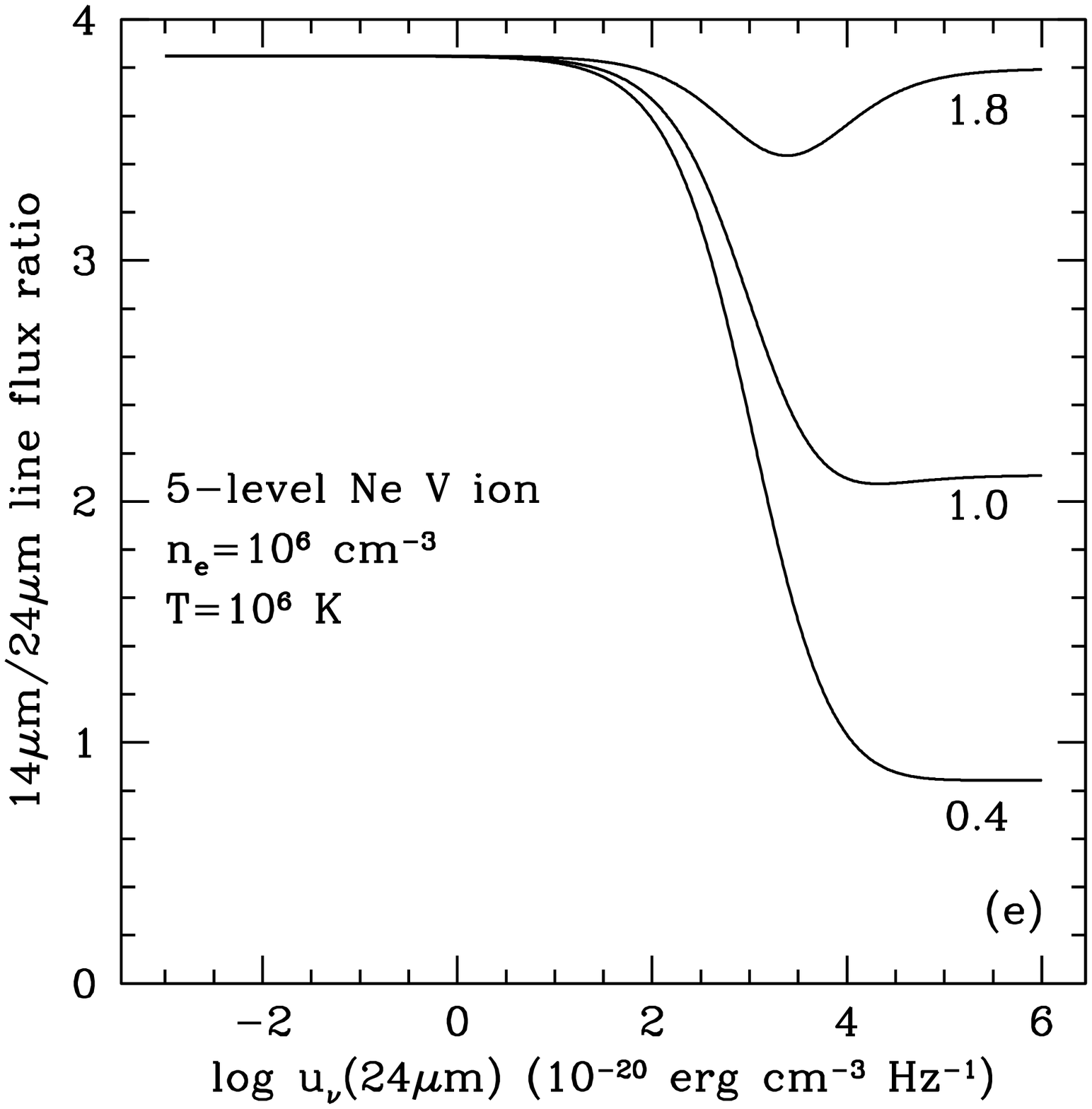}}\\
\end{center}
\caption[]{The [NeV] line ratio as a function of the specific energy density at 24$\mu$m, $u_{\nu}$(24$\mu$m), for temperatures, T = 10$^4$, 10$^5$, 10$^6$ K, for 14$\mu$m/24$\mu$m continuum ratios of 0.4, 1.0, and 1.8, and finally for electron densites, n$_e$ = 10$^2$, 10$^3$, 10$^4$, 10$^5$, 10$^6$ cm$^{-3}$.}
\end{figure*}
									
In Section 5.1, we suggest that the [NeV]-emitting region may lie within the 
torus.  In this case, absorption and stimulated emission within the ground
term are probably important.  For the high-luminosity objects, these may
even dominate over collisional excitation and de-excitation.  At these 
central locations, gas temperatures $T \sim 10^6 \K$ may be natural
(Ferland et al.~2002).  Relatively high densities may also be expected, 
in which case the infrared continuum may not appreciably depress the line
ratio (see Figure 4d).

Adopting the Mathews \& Ferland (1987) spectrum and $T \approx 10^6 \K$,
the ionization parameter $U \equiv n_{\gamma}/n_e \sim 10^{-3}$ in order
for a substantial fraction of the Ne to be NeV; $n_{\gamma}$ is the 
number density of H-ionizing photons.  For this spectrum, 
$n_{\gamma} \approx 1.7 \times 10^3 L_{\rm bol, \, 46} \, 
r_{\rm Ne, \, 100}^{-2} \cm^{-3}$, where 
$r_{\rm Ne, \, 100} = r_{\rm Ne}/ 100 \pc$.
If $r_{\rm Ne} = 1 \pc$, then either (1) $n_e \sim 10^{10} L_{\rm bol, \, 46}
\cm^{-3}$ or (2) the nuclear continuum is filtered through a far-UV/X-ray-absorbing
medium before reaching the [NeV]-emitting region.

If absorption and stimulated emission are indeed relevant processes in [NeV] line production, we might expect a relationship between the [NeV] line flux ratio and the 24 $\mu$m continuum luminosity that is consistent with one of the curves shown in Figures 4a through 4e.  In Figure 5 we plot this relationship for the [NeV] emitting galaxies in our sample.  As can be seen in Figure 5, we find no relationship between the [NeV] line flux ratio and the 24$\mu$m continuum luminosity for our sample of galaxies.  The Spearman rank correlation coefficient for this plot is -0.01 (probability of chance correlation = 0.95), indicating no correlation.  As a result, as can be seen from Figures 4a and 4b, stimulated emission and absorption at low densities can be ruled out as possible scenarios because the scatter plot shown in Figure 5 does not follow the model predictions.  We note that the location of the [NeV]-emitting region relative to the source of the 24$\mu$m continuum emission is uniform among the galaxies in the sample.  Variations in the location might obscure any correlation in these plots.   Figures 4c and 4d reveal that, for some values of n$_e$, T, and u$_{\nu}$(14$\mu$m)/u$_{\nu}$(24$\mu$m), the line ratio is very insensitive to the value of u$_{\nu}$(24$\mu$m).  In these cases, the line ratio remains above $\sim$ 0.8.  Thus, although absorption and stimulated emission may be contributing processes to [NeV] production, {\it another mechanism is required to explain the low ($<$0.8) [NeV] line flux ratios in our sample.}

\begin{figure}[htbp]
\begin{center}
{\includegraphics[width=9cm]{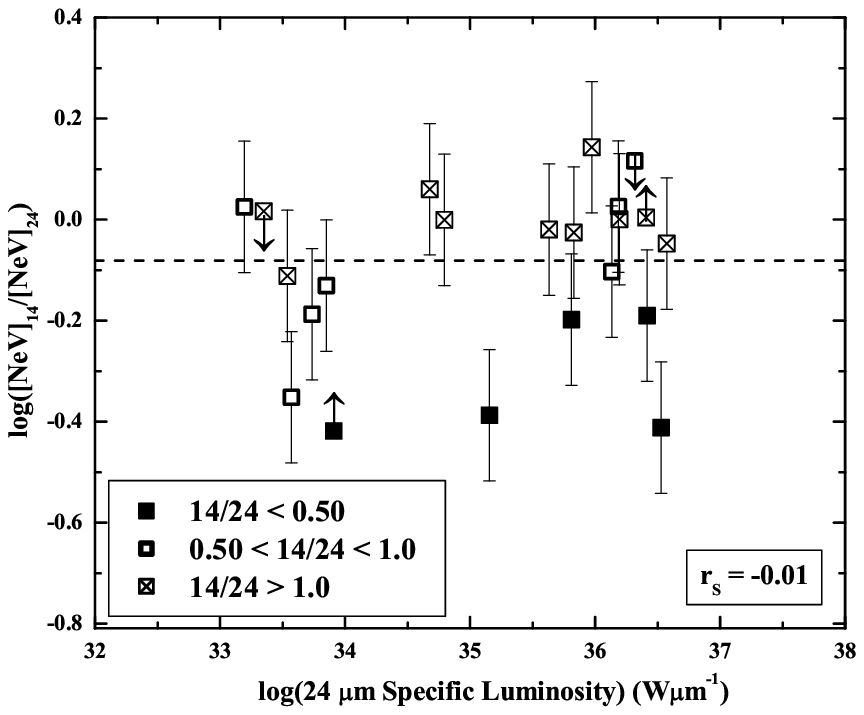}}\\
\end{center}
\caption[]{The observed [NeV] line ratio as a function of the 24$\mu$m specific luminosity for our sample of galaxies.  The error bars shown here represent the calibration uncertainties on the [NeV] line flux ratio as in Figure 3.  The symbol type indicates the 14$\mu$m/24$\mu$m continuum ratio.}
\end{figure}

{\bf Computed Quantities:}  Finally, it is possible that there is significant error in the adopted collisional rate coefficients.   The accuracy of collisional strengths of infrared atomic transitions has been a longstanding question.  We adopt the collisional rate coefficients from the state of the art IRON project (Hummer et al. 1993) which produced the most up-to-date and accurate collision strengths for a large database of atomic transitions.  While these calculations have been questioned based on recent ISO observations of nebulae (Clegg et al. 1987, Oliva et al. 1996, Rubin et al. 2002, Rubin 2004), it is likely that the discrepancies between the observational and theoretical values can be explained by inaccuracies in the fluxes employed (van Hoof et al. 2000).  Uncertainties in the collisional rate coefficients for the [NeV] transitions are unlikely to exceed 30\% (van Hoof et al. 2000).  It is therefore unlikely that the low critical densities implied by our data can be attributed to uncertainties in the theoretical values of the [NeV] collisional strengths.


\begin{table*}
\fontsize{8pt}{9pt}\selectfont
\begin{center}
\begin{tabular}{lccccccccc}
\multicolumn{10}{l}{{\bf Table 2: NeV Line Fluxes and Derived Extinction}}  \\ 
 \hline

\multicolumn{1}{c}{Galaxy} & \multicolumn{1}{c}{[NeV]} & [NeV] & [NeV] & [NeV] & [NeV] & A$_V$  & A$_V$ & A$_V$ & A$_V$\\

\multicolumn{1}{c}{Source} & \multicolumn{1}{c}{14.32} & 14.32 & 24.32 & 24.32 &Ratio & Ratio to& Ratio to & Ratio to& Ratio to \\

\multicolumn{1}{c}{} & \multicolumn{1}{c}{SH} & ISO & LH & ISO & & LDL (D\&L) & HDL (D\&L) & LDL (C\&T) & HDL (C\&T)  \\
\multicolumn{1}{c}{(1)} & \multicolumn{1}{c}{(2)} & (3) & (4) & (5) & (6) & (7) & (8) & (9) & (10)\\ 

\hline
\multicolumn{10}{l}{{\it Seyferts}}\\
\hline

NGC4151 & 7.77$^a$ & 5.50$^c$ & 6.77$^a$ & 5.60$^c$ & 1.15$^{+0.41}_{-0.30}$ & $\cdots$ & 126 & $\cdots$ & 1686\\

NGC1365 & 2.20$\pm$0.06 & 2.50$^c$ & 5.36$\pm$0.06 & 3.90$^c$ & 0.41$^{+0.14}_{-0.11}$ & 45 & 189 & 570 & 2519\\

NGC1097 & $<$0.05 & $\cdots$ & $<$0.18 & $\cdots$ & $\cdots$ & $\cdots$ & $\cdots$ & $\cdots$ & $\cdots$\\

NGC7469 & 1.16$^a$ & $<$1.50$^c$ & 1.47$^a$ & 0.63$^{c*}$ & 0.79$^{+0.28}_{-0.21}$ & 3 & 149 & 41 & 1990\\

NGC4945 & 0.28$\pm$0.03 & $<$0.50$^d$ & $<$0.75 & $\cdots$ & $>$0.38 & $\cdots$ & $\cdots$ & $\cdots$ & $\cdots$\\

Circinus & 23.94$\pm$0.61 & 31.70$^c$ & 24.00$\pm$3.90 & 21.80$^c$ & 1.00$^{+0.35}_{-0.26}$ & $\cdots$ & 135 & $\cdots$ & 1799\\

Mrk 231 & $<$0.44$^a$ & $<$1.50$^e$ & $<$0.69$^a$ & $\cdots$ & $\cdots$ & $\cdots$ & $\cdots$ & $\cdots$ & $\cdots$\\

Mrk3 & 6.45$^a$ & 4.60$^c$ & 6.75$^a$ & 3.40$^c$ & 0.96$^{+0.34}_{-0.25}$ & $\cdots$ & 138 & $\cdots$ & 1835\\

Cen A & 2.32$^a$ & 2.70$^c$ & 2.99$^a$ & 2.00$^c$ & 0.77$^{+0.27}_{-0.20}$ & 4 & 150 & 56 & 2005\\

Mrk463 & 1.83$^b$ & 1.40$^c$ & 2.04$^b$ & $\cdots$ & 0.90$^{+0.32}_{-0.23}$ & $\cdots$ & 141 & $\cdots$ & 1886\\

NGC 4826 & $\cdots$ & $\cdots$ & $\cdots$ & $\cdots$ & $\cdots$ & $\cdots$ & $\cdots$ & $\cdots$ & $\cdots$\\

NGC 4725 & $<$0.09 & $\cdots$ & 0.09$\pm$0.03 & $\cdots$ & $<$1.04 & $\cdots$ & $\cdots$ & $\cdots$ & $\cdots$\\

1 ZW 1 & $<$0.11$^a$ & 0.27$^c$ & $<$0.10$^a$ & $\cdots$ & $\cdots$ & $\cdots$ & $\cdots$ & $\cdots$ & $\cdots$\\

NGC 5033 & 0.07$\pm$0.02 & $\cdots$ & 0.11$\pm$0.02 & $\cdots$ & 0.65$^{+0.23}_{-0.17}$ & 16 & 161 & 198 & 2146\\

NGC1566 & 0.16$\pm$0.05 & $\cdots$ & 0.22$\pm$0.04 & $\cdots$ & 0.74$^{+0.26}_{-0.19}$ & 7 & 153 & 92 & 2041\\

NGC 2841 & $<$0.04 & $\cdots$ & $<$0.03 & $\cdots$ & $\cdots$ & $\cdots$ & $\cdots$ & $\cdots$ & $\cdots$\\

NGC 7213 & $<$0.04 & $\cdots$ & $<$0.09 & $\cdots$ & $\cdots$ & $\cdots$ & $\cdots$ & $\cdots$ & $\cdots$\\

\hline
\multicolumn{10}{l}{{\it LINERs}}\\
\hline

NGC4579 & $<$0.06 & $\cdots$ & $<$0.03 & $\cdots$ & $\cdots$ & $\cdots$ & $\cdots$ & $\cdots$ & $\cdots$\\

NGC3031 & $<$0.06 & $\cdots$ & $<$0.04 & $\cdots$ & $\cdots$ & $\cdots$ & $\cdots$ & $\cdots$ & $\cdots$\\

NGC6240 & 0.51$^b$ & $<$1.00$^e$ & $<$0.39$^b$ & $\cdots$ & $<$1.31 & $\cdots$ & $\cdots$ & $\cdots$ & $\cdots$\\

NGC5194 & 0.41$\pm$0.04 & $<$0.20$^c$ & 0.39$\pm$0.09 & $\cdots$ & 1.06$^{+0.37}_{-0.28}$ & $\cdots$ & 131 & $\cdots$ & 1751\\

MRK266$^{**}$ & 0.21$\pm$0.02 & 0.50$^f$ & 1.19$\pm$0.06 & $\cdots$ & 0.18$^{+0.06}_{-0.05}$ & 100 & 240 & 1254 & 3203\\

NGC7552 & $<$0.11 & $\cdots$ & $<$0.83 & $\cdots$ & $\cdots$ & $\cdots$ & $\cdots$ & $\cdots$ & $\cdots$\\

NGC 4552 & $<$0.06 & $\cdots$ & $<$0.07 & $\cdots$ & $\cdots$ & $\cdots$ & $\cdots$ & $\cdots$ & $\cdots$\\

NGC 3079 & $<$0.07$^a$ & $\cdots$ & $<$0.14$^a$ & $\cdots$ & $\cdots$ & $\cdots$ & $\cdots$ & $\cdots$ & $\cdots$\\

NGC 1614 & $<$0.28 & $\cdots$ & $<$1.49 & $\cdots$ & $\cdots$ & $\cdots$ & $\cdots$ & $\cdots$ & $\cdots$\\

NGC 3628 & $<$0.06 & $\cdots$ & $<$0.34 & $\cdots$ & $\cdots$ & $\cdots$ & $\cdots$ & $\cdots$ & $\cdots$\\

NGC 2623 & 0.30$\pm$0.04 & $\cdots$ & 0.47$\pm$0.07 & $\cdots$ & 0.63$^{+0.22}_{-0.14}$ & 17 & 163 & 218 & 2167\\

IRAS23128$\cdots$ & 0.22$\pm$0.02 & $<$0.40$^e$ & 0.34$\pm$0.10 & $\cdots$ & 0.65$^{+0.23}_{-0.22}$ & 16 & 161 & 203 & 2152\\

MRK273 & 1.06$\pm$0.05 & 0.82$^e$ & 2.74$\pm$0.19 & $\cdots$ & 0.39$^{+0.14}_{-0.10}$ & 49 & 192 & 617 & 2565\\

IRAS20551$\cdots$ & $<$0.06 & $<$0.25$^e$ & $<$0.25 & $\cdots$ & $\cdots$ & $\cdots$ & $\cdots$ & $\cdots$ & $\cdots$\\

NGC3627 & 0.08$\pm$0.01 & $\cdots$ & 0.19$\pm$0.05 & $\cdots$ & 0.45$^{+0.16}_{-0.12}$ & 40 & 184 & 504 & 2453\\

UGC05101 & 0.52$^b$ & $<$1.50$^e$ & 0.49$^b$ & $\cdots$ & 1.06$^{+0.37}_{-0.28}$ & $\cdots$ & 131 & $\cdots$ & 1750\\

NGC4125 & $<$0.03 & $\cdots$ & $<$0.07 & $\cdots$ & $\cdots$ & $\cdots$ & $\cdots$ & $\cdots$ & $\cdots$\\

NGC 4594 & $<$0.03 & $\cdots$ & $<$0.04 & $\cdots$ & $\cdots$ & $\cdots$ & $\cdots$ & $\cdots$ & $\cdots$\\

\hline
\multicolumn{10}{l}{{\it Quasars}}\\
\hline

PG1351$\cdots$ & $<$0.04 & $\cdots$ & $<$0.07 & $\cdots$ & $\cdots$ & $\cdots$ & $\cdots$ & $\cdots$ & $\cdots$\\

PG1211$\cdots$ & 0.04$\pm$0.007 & $\cdots$ & $<$0.04 & $\cdots$ & $>$1.01 & $\cdots$ & $\cdots$ & $\cdots$ & $\cdots$\\

PG1119$\cdots$ & 0.30$\pm$0.06 & $\cdots$ & 0.22$\pm$0.02 & $\cdots$ & 1.39$^{+0.49}_{-0.36}$ & $\cdots$ & 115 & $\cdots$ & 1531\\

PG2130$\cdots$ & 0.42$\pm$0.03 & $\cdots$ & 0.42$\pm$0.05 & $\cdots$ & 1.00$^{+0.35}_{-0.26}$ & $\cdots$ & 135 & $\cdots$ & 1798\\

PG0804$\cdots$ & $<$0.06 & $\cdots$ & $<$0.07 & $\cdots$ & $\cdots$ & $\cdots$ & $\cdots$ & $\cdots$ & $\cdots$\\

PG1501$\cdots$ & 0.78$\pm$0.02 & $\cdots$ & 0.83$\pm$0.02 & $\cdots$ & 0.94$^{+0.33}_{-0.25}$ & $\cdots$ & 138 & $\cdots$ & 1846\\

\hline
\end{tabular}
\end{center}
{\scriptsize{\bf Columns Explanation:} Col(1):Common Source Names; 
Col(2):  14.32 $\mu$m [NeV] line flux and statistical error in units of 10$^{-20}$ W cm$^{-2}$ from {\it Spitzer};
Col(3):  14.32 $\mu$m [NeV] line flux and statistical error in units of 10$^{-20}$ W cm$^{-2}$ from {\it ISO};
Col(4):  24.31 $\mu$m [NeV] line flux and statistical error in units of 10$^{-20}$ W cm$^{-2}$ from {\it Spitzer};
Col(5):  24.32 $\mu$m [NeV] line flux and statistical error in units of 10$^{-20}$ W cm$^{-2}$ from {\it ISO};  
Col(6): [NeV] Line Ratio used in plots and calculations; 
Col(7): Extinction required to bring ratios below the low-density limit (LDL) up to the LDL, calculated using the Draine (1989) extinction curve amended by the more recent {\it ISO} SWS extinction curve toward the Galactic center for 2.5-10$\mu$m (Lutz et al. 1996);
Col(8): Extinction required to bring ratios below the low-density limit (LDL) up to the high-density limit (HDL), calculated using the Draine (1989) extinction curve amended by the more recent {\it ISO} SWS extinction curve toward the Galactic center for 2.5-10$\mu$m (Lutz et al. 1996),
Col(9): Extinction required to bring ratios below the low-density limit (LDL) up to the LDL, calculated using the Chiar \& Tielens (2006) extinction curve for the local ISM,
Col(10): Extinction required to bring ratios below the low-density limit (LDL) up to the high-density limit (HDL), calculated using the Chiar \& Tielens (2006) extinction curve for the local ISM.$*$ Sturm et al. 2002 find that the [NeV] 24$\mu$m detection for NGC 7469 is a questionable one, $**$ As discussed in detail in Section 6.2, Mrk 266 is the only galaxy in our sample where we find that aperture variation may affect the observed [NeV] line flux ratio.  For this reason it has been excluded from relevent plots and  calculations.}
{\scriptsize{\bf References for Table 2:}$^a$ Weedman et al. 2005, $^b$ Armus et al. 2004 \& 2006, $^c$ Sturm et al. 2002, $^d$ Verma et al. 2003, $^e$ Genzel et al. 1998, $^f$ Prieto \& Viegas 2000}
\end{table*}


\begin{table*}
\fontsize{8pt}{9pt}\selectfont
\begin{center}
\begin{tabular}{lcccccccccc}
\multicolumn{11}{l}{{\bf Table 3: SIII Line Fluxes and Derived Extinction}}  \\\hline

\multicolumn{1}{c}{Galaxy} & \multicolumn{1}{c}{[SIII]} & [SIII] & [SIII] & [SIII] & [SIII] & A$_v$  & A$_v$ & A$_v$ & A$_v$ & PAH$_{6.2}$\\

\multicolumn{1}{c}{Source} & \multicolumn{1}{c}{18.71} & 18.71 & 33.48 & 33.48 &Ratio & Ratio to & Ratio to & Ratio to & Ratio to & SL2 \\

\multicolumn{1}{c}{} &  \multicolumn{1}{c}{SH} & ISO & LH & ISO & & LDL (D\&L) & HDL (D\&L) & LDL (C\&T) & HDL (C\&T) & \\

\multicolumn{1}{c}{(1)} & \multicolumn{1}{c}{(2)} & (3) & (4) & (5) & (6) & (7) & (8)& (9) & (10) & (11)\\ 

\hline
\multicolumn{11}{l}{{\it Seyferts}}\\
\hline 

NGC4151 & 7.50$^a$ & 5.40$^c$ & 6.57$^a$ & 8.10$^c$ & 1.14$^{+0.40}_{-0.30}$ & $\cdots$ & $\cdots$ & $\cdots$ & $\cdots$ & 168.1\\

NGC1365 & 5.73$\pm$0.05 & 13.50$^c$ & 27.20$\pm$0.38 & 36.10$^c$ & 0.21$^{+0.07}_{-0.05}$ & $\cdots$ & $\cdots$ & $\cdots$ & $\cdots$ & 132.3\\

NGC1097 & 2.18$\pm$0.02 & $\cdots$ & 11.40$\pm$0.23 & $\cdots$ & 0.19$^{+0.07}_{-0.05}$ & $\cdots$ & $\cdots$ & $\cdots$ & $\cdots$ & 151.7\\

NGC7469 & 7.70$^a$ & 9.20$^c$ & 9.80$^a$ & 10.40$^c$ & 0.79$^{+0.28}_{-0.20}$ & $\cdots$ & $\cdots$ & $\cdots$ & $\cdots$ & 415.0\\

NGC4945 & 3.18$\pm$0.03 & 6.30$^d$ & 38.70$\pm$1.80 & 51.40$^d$ & 0.08$^{+0.03}_{-0.02}$ & $\cdots$ & $\cdots$ & $\cdots$ & $\cdots$ & 671.7\\

Circinus & 19.10$\pm$0.70 & 35.20$^c$ & 56.30$\pm$3.31 & 93.20$^c$ & 0.37$^{+0.13}_{-0.10}$ & $\cdots$ & $\cdots$ & $\cdots$ & $\cdots$ & 1018.4\\

Mrk 231 & $<$0.47$^a$ & $<$3.00$^e$ & $<$2.30$^a$ & $<$3.00$^e$ & $\cdots$ & $\cdots$ & $\cdots$ & $\cdots$ & $\cdots$ & 175.6\\

Mrk3 & 5.55$^a$ & $\cdots$ & 5.25$^a$ & $\cdots$ & 1.06$^{+0.37}_{-0.28}$ & $\cdots$ & 83 & $\cdots$ & 72 & 18.4\\

Cen A & 4.54$^a$ & 6.40$^c$ & 14.80$^a$ & 22.30$^c$ & 0.31$^{+0.11}_{-0.08}$ & $\cdots$ & $\cdots$ & $\cdots$ & $\cdots$ & 220.8\\

Mrk463 & 1.50$^b$ & $<$0.80$^c$ & 1.35$^b$ & 1.20$^c$ & 1.11$^{+0.39}_{-0.29}$ & $\cdots$ & 81 & $\cdots$ & 70 & 55.4\\

NGC 4826 & 3.39$\pm$0.03$^f$ & $\cdots$ & 4.61$\pm$0.08$^f$ & $\cdots$ & 0.74$^{+0.26}_{-0.19}$ & $\cdots$ & 96 & $\cdots$ & 82 & 66.2\\

NGC 4725 & 0.02$\pm$0.02$^f$ & $\cdots$ & 0.11$\pm$0.02$^f$ & $\cdots$ & 0.23$^{+0.08}_{-0.06}$ & 23 & 135 & 20 & 116 & 3.7\\

1 ZW 1 & $<$0.11$^a$ & $<$0.50$^c$ & $<$0.18$^a$ & $<$1.00$^c$ & $\cdots$ & $\cdots$ & $\cdots$ & $\cdots$ & $\cdots$ & 22.2\\

NGC 5033 & 0.85$\pm$0.11 & $\cdots$ & 2.42$\pm$0.10 & $\cdots$ & 0.35$^{+0.12}_{-0.09}$ & $\cdots$ & $\cdots$ & $\cdots$ & $\cdots$ & 33.9\\

NGC1566 & 0.55$\pm$0.05$^f$ & $\cdots$ & 0.55$\pm$0.06$^f$ & $\cdots$ & 1.00$^{+0.35}_{-0.26}$ & $\cdots$ & 85 & $\cdots$ & 73 & 61.0\\

NGC 2841 & 0.22$\pm$0.04$^f$ & $\cdots$ & 0.29$\pm$0.03$^f$ & $\cdots$ & 0.75$^{+0.26}_{-0.20}$ & $\cdots$ & 95 & $\cdots$ & 82 & 10.9\\

NGC 7213 & 0.47$\pm$0.05 & $\cdots$ & 0.59$\pm$0.06 & $\cdots$ & 0.80$^{+0.28}_{-0.21}$ & $\cdots$ & $\cdots$ & $\cdots$ & $\cdots$ & 28.7\\

\hline
\multicolumn{11}{l}{{\it LINERs}}\\
\hline
NGC4579 & 0.32$\pm$0.06$^f$ & $<$0.78$^g$ & 0.24$\pm$0.03$^f$ & $<$1.20$^g$ & 1.33$^{+0.47}_{-0.35}$ & $\cdots$ & 75 & $\cdots$ & 65 & 17.3\\

NGC3031 & 0.61$\pm$0.03 & $\cdots$ & 0.09$\pm$0.09 & $\cdots$ & 0.67$^{+0.24}_{-0.18}$ & $\cdots$ & $\cdots$ & $\cdots$ & $\cdots$ & 23.4\\

NGC6240 & 1.99$^b$ & $<$4.00$^e$ & 2.63$^b$ & 4.50$^e$ & 0.76$^{+0.27}_{-0.20}$ & $\cdots$ & 95 & $\cdots$ & 81 & 399$^b$\\

NGC5194 & 1.06$\pm$0.05$^f$ & 1.00$^d$ & 1.48$\pm$0.03$^f$ & 4.60$^d$ & 0.72$^{+0.25}_{-0.19}$ & $\cdots$ & 96 & $\cdots$ & 83 & 26.4\\

MRK266$^{**}$ & 1.00$\pm$0.13 & $\cdots$ & 4.65$\pm$0.09 & $\cdots$ & 0.21$^{+0.08}_{-0.06}$ & 25 & 138 & 22 & 118 & 23.1\\

NGC7552 & 17.11$\pm$0.08$^f$ & 24.60$^d$ & 13.38$\pm$0.41$^f$ & 41.10$^d$ & 1.28$^{+0.45}_{-0.33}$ & $\cdots$ & 77 & $\cdots$ & 66 & 872.0\\

NGC 4552 & 0.07$\pm$0.03$^f$ & $\cdots$ & 0.06$\pm$0.02$^f$ & $\cdots$ & 1.29$^{+0.45}_{-0.34}$ & $\cdots$ & 76 & $\cdots$ & 66 & 21.5\\

NGC 3079 & 1.25$^a$ & 6.80$^g$ & 6.08$^a$ & 6.60$^g$ & 0.21$^{+0.07}_{-0.05}$ & $\cdots$ & $\cdots$ & $\cdots$ & $\cdots$ & 620.3\\

NGC 1614 & 9.63$\pm$0.27 & $\cdots$ & 11.60$\pm$0.43 & $\cdots$ & 0.83$^{+0.29}_{-0.15}$ & $\cdots$ & 91 & $\cdots$ & 79 & 508.5\\

NGC 3628 & 2.14$\pm$0.03 & $\cdots$ & 15.80$\pm$0.33 & $\cdots$ & 0.14$^{+0.05}_{-0.04}$ & $\cdots$ & $\cdots$ & $\cdots$ & $\cdots$ & 430.4\\

NGC 2623 & 0.88$\pm$0.05 & $\cdots$ & 3.16$\pm$0.20 & $\cdots$ & 0.28$^{+0.10}_{-0.07}$ & 16 & 129 & 14 & 111 & 128.6\\

IRAS23128$\cdots$ & 2.62$\pm$0.12 & 0.89$^e$ & 2.11$\pm$0.18 & 2.80$^e$ & 1.24$^{+0.44}_{-0.32}$ & $\cdots$ & 78 & $\cdots$ & 67 & 90.1\\

MRK273 & 1.24$\pm$0.07 & $<$0.82$^e$ & 3.88$\pm$0.40 & 2.30$^e$ & 0.32$^{+0.11}_{-0.08}$ & 12 & 124 & 10 & 107 & 69.2\\

IRAS20551$\cdots$ & 0.66$\pm$0.06 & 0.30$^e$ & 1.18$\pm$0.13 & 1.40$^e$ & 0.56$^{+0.20}_{-0.15}$ & $\cdots$ & 105 & $\cdots$ & 90 & 38.3\\

NGC3627 & 0.38$\pm$0.03$^f$ & $\cdots$ & 0.57$\pm$0.09$^f$ & $\cdots$ & 0.67$^{+0.24}_{-0.18}$ & $\cdots$ & 99 & $\cdots$ & 85 & 153.8\\

UGC05101 & 0.98$^b$ & $<$1.40$^e$ & 1.30$^b$ & 2.50$^e$ & 0.75$^{+0.27}_{-0.20}$ & $\cdots$ & 95 & $\cdots$ & 82 & 190$^b$\\

NGC4125 & $\cdots$$^f$ & $\cdots$ & 0.06$\pm$0.05$^f$ & $\cdots$ & $\cdots$ & $\cdots$ & $\cdots$ & $\cdots$ & $\cdots$ & 14.9\\

NGC 4594 & 0.39$\pm$0.03 & $\cdots$ & 1.24$\pm$0.13 & $\cdots$ & 0.32$^{+0.11}_{-0.08}$ & $\cdots$ & $\cdots$ & $\cdots$ & $\cdots$ & 14.2\\

\hline
\multicolumn{11}{l}{{\it Quasars}}\\
\hline

PG1351$\cdots$ & 0.34$\pm$0.06 & $\cdots$ & $<$0.13 & $\cdots$ & $>$2.70 & $\cdots$ & $\cdots$ & $\cdots$ & $\cdots$ & 29.6\\

PG1211$\cdots$ & $<$0.06 & $\cdots$ & $<$0.08 & $\cdots$ & $\cdots$ & $\cdots$ & $\cdots$ & $\cdots$ & $\cdots$ & 25.8\\

PG1119$\cdots$ & $<$0.13 & $\cdots$ & 0.19$\pm$0.06 & $\cdots$ & $<$0.71 & $\cdots$ & $\cdots$ & $\cdots$ & $\cdots$ & 8.7\\

PG2130$\cdots$ & $<$0.19 & $\cdots$ & 0.34$\pm$0.06 & $\cdots$ & $<$0.55 & $\cdots$ & $\cdots$ & $\cdots$ & $\cdots$ & 26.9\\

PG0804$\cdots$ & $<$0.06 & $\cdots$ & $<$0.21 & $\cdots$ & $\cdots$ & $\cdots$ & $\cdots$ & $\cdots$ & $\cdots$ & 27.4\\

PG1501$\cdots$ & 0.67$\pm$0.15 & $\cdots$ & 0.41$\pm$0.05 & $\cdots$ & 1.64$^{+0.58}_{-0.43}$ & $\cdots$ & 68 & $\cdots$ & 59 & 19.2\\

\hline
\end{tabular}
\end{center}
{\scriptsize{\bf Columns Explanation:} Col(1):Common Source Names; 
Col(2):  18.71 $\mu$m [SIII] line flux and statistical error in units of 10$^{-20}$ W cm$^{-2}$ from {\it Spitzer};
Col(3):  18.71 $\mu$m [SIII] line flux and statistical error in units of 10$^{-20}$ W cm$^{-2}$ from {\it ISO};
Col(4):  33.48 $\mu$m [SIII] line flux and statistical error in units of 10$^{-20}$ W cm$^{-2}$ from {\it Spitzer};
Col(5):  33.48 $\mu$m [SIII] line flux and statistical error in units of 10$^{-20}$ W cm$^{-2}$ from {\it ISO};
Col(6):[SIII] line flux ratio used for plots and calculations;
Col(7): Extinction required to bring ratios below the low-density limit (LDL) up to the LDL, calculated using the Draine (1989) extinction curve amended by the more recent {\it ISO} SWS extinction curve toward the Galactic center for 2.5-10$\mu$ (Lutz et al. 1996) for those galaxies with distances greater than 55 Mpc that are not effected by aperture variations, 
Col(8): Extinction required to bring ratios below the low-density limit (LDL) up to the high-density limit (HDL), calculated using the Draine (1989) extinction curve amended by the more recent {\it ISO} SWS extinction curve toward the Galactic center for 2.5-10$\mu$ (Lutz et al. 1996) for those galaxies with distances greater than 55 Mpc that are not effected by aperture variations,
Col(9): Extinction required to bring ratios below the low-density limit (LDL) up to the LDL, calculated using the Chiar \& Tielens (2006) extinction curve for the Galactic Center for those galaxies with distances greater than 55 Mpc that are not effected by aperture variations,
Col(10): Extinction required to bring ratios below the low-density limit (LDL) up to the high-density limit (HDL), calculated using the Chiar \& Tielens (2006) extinction curve for the Galactic Center for those galaxies with distances greater than 55 Mpc that are not effected by aperture variations,
Col(11): 6.2 $\mu$m PAH line flux in units of 10$^{-21}$ W cm$^{-2}$ 
$**$ The [SIII] ratio for Mrk 266 is known to be affected by aperture variations(See Section 6.2).  For this reason it has been excluded from relevent plots and calculations.}
{\scriptsize{\bf References for Table 3:}$^a$ Weedman et al. 2005, $^b$ Armus et al. 2004 \& 2006, $^c$ Sturm et al. 2002, $^d$ Verma et al. 2003, $^e$ Genzel et al. 1998,  $^f$ Dale et al. 2006, $^g$ Satyapal et al. 2004. }
\end{table*}
\section{Extinction Effects of the Torus and AGN Unification}  

Although low electron densities, high gas temperatures, and/or high infrared radiation densities may play a role in lowering the [NeV] line flux ratio, we argue that differential infrared extinction to the [NeV] emitting region due to dust in the obscuring torus is responsible for the low line ratios in at least some AGN. Clearly, this requires that there is significant extinction at mid-IR wavelengths, and specifically toward the [NeV]-emitting regions.  Is this reasonable?  If there is significant extinction, it is possible that:  1) the [NeV]-emitting region originates much closer to the central source than previously recognized, close enough to be extinguished by the central torus in some galaxies, 2) the [NeV]-emitting portion of the NLR is obscured by dust in the host galaxy or in the NLR itself, or 3) some combination of these scenarios.  We explore these possibilities in the following analysis.  

\subsection{The [NeV] originates in gas interior to the central torus.}  
In the conventional picture of an AGN, the broad line region (BLR) is thought to exist within a small region interior to a dusty molecular torus while the NLR originates further out.  This of course is the paradigm invoked to explain the Type 1/Type 2 dichotomy.  However there have been multiple optical spectroscopic studies that contradict the assumption that the observational properties of the NLR are not dependent on the viewing angle and the inclination of the system, suggesting that some of the narrow emission lines originate in gas interior to the torus.  For instance, Shuder and Osterbrock (1981) and Cohen (1983) showed that narrow high ionization forbidden lines such as [Fe VII] $\lambda$ 6374 (requiring photons with energies $\geq$ 100eV to ionize) are stronger relative to the low ionization lines in Seyfert 1 galaxies (including intermediate Seyferts, 1.2, 1.5 etc.) than in Seyfert 2 galaxies, suggesting that some of the emission is obscured by the torus.  In addition, [FeX] $\lambda$ 6374 and [NeV] $\lambda$ 3426 have also been shown to be less luminous in Type 2 objects than in Type 1 objects (Murayama \& Taniguchi, 1998a; Schmitt 1998, Nagao et al. 2000, 2001a, 2001b, 2003, Tran et al. 2000, see also Jackson and Browne (1990) for narrow line radio galaxies and quasars.)  These findings may imply that the emission lines of species with the highest ionization potentials originate closer to the AGN than those of  lower ionization species such as [OII]$\lambda$3727, [SII]$\lambda$$\lambda$6716, 6731, [OI] $\lambda$6300 etc. and therefore may be partially obscured by the central torus.

 If there is considerable extinction to the line-emitting regions due to the torus, one may expect the mid-infrared continuum to be similarly obscured.  To test this scenario we divided our sample into Type 1 or Type 2 objects based on the presence or absence of broad (full width at half max (FWHM) exceeding 1000 km s$^{-1}$) Balmer emission lines in the optical spectrum.  The spectral classification for the [NeV]-emitting galaxies is given in Table 1.  In Figure 6a, we plot the [NeV] 14$\mu$m/24$\mu$m line flux ratio versus the 14$\mu$m/24$\mu$m continuum ratio of the [NeV] emitting galaxies in our sample.  Assuming there is no correlation between the electron density and the continuum shape, a correlation between the line flux and continuum ratios would suggest that the mid-IR extinction associated with the torus (such as that found by Clavel et al. 2000) affects the observed line flux ratios.  As can be seen, there is a correlation between the line and continuum ratios for galaxies with [NeV] emission.  Moreover we note that the 3 nuclei with ratios significantly below the LDL are all Type 2 AGNs, while the 2 that lie significantly above this limit are Type 1 AGNs, suggestive that the extinction of the [NeV]-emitting region in Type 2 AGNs may be due to the torus. We note that the error bars displayed in Figure 6 are based on a conservative estimate (15\%) of the absolute calibration error on the flux (see Section 3).  Moreover, we have adopted the most conservative approach in propagating the error (see Section 4) for each line ratio.  We further note that two of the three nuclei with ratios below the LDL were also observed in high-accuracy peak-up mode, resulting in a pointing accuracy on the continuum for these galaxies of 0.4''.  The third galaxy, NGC 3627, was observed in high resolution mapping mode over 15'' X 22''.  We extracted the spectra and found that the full map and the single slit fluxes agree to within 10\%.  Thus pointing errors do not appear to be responsible for the low ratios in these galaxies.   Finally we find that the ratios for the all galaxies except Mrk 266 are not sensitive to the line-fitting or flux extraction methods that we have employed.

\begin{figure*}[htbp]
\begin{center}
{\includegraphics[width=9cm]{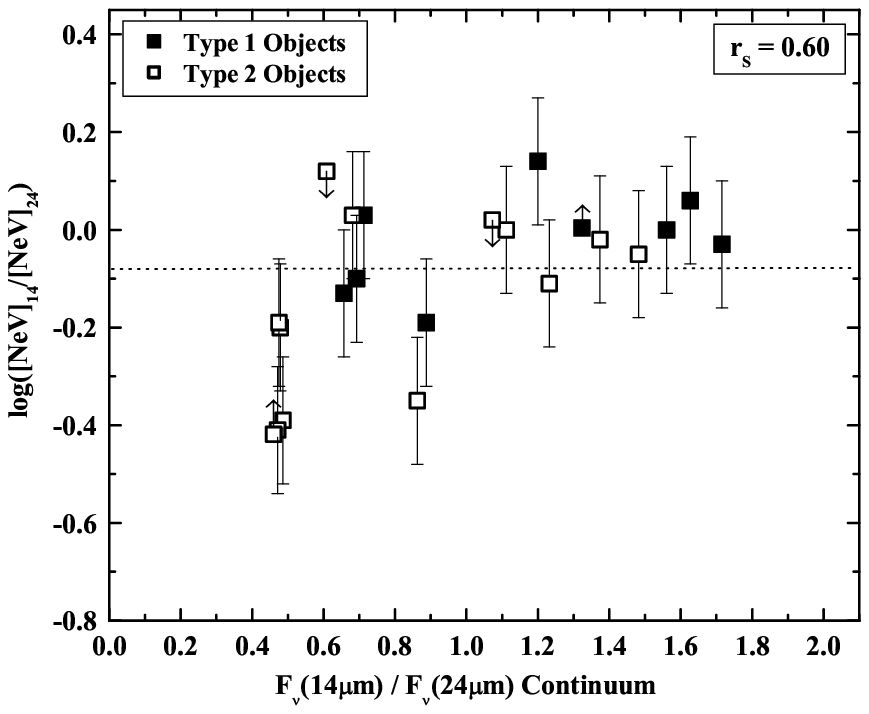}}{\includegraphics[width=9cm]{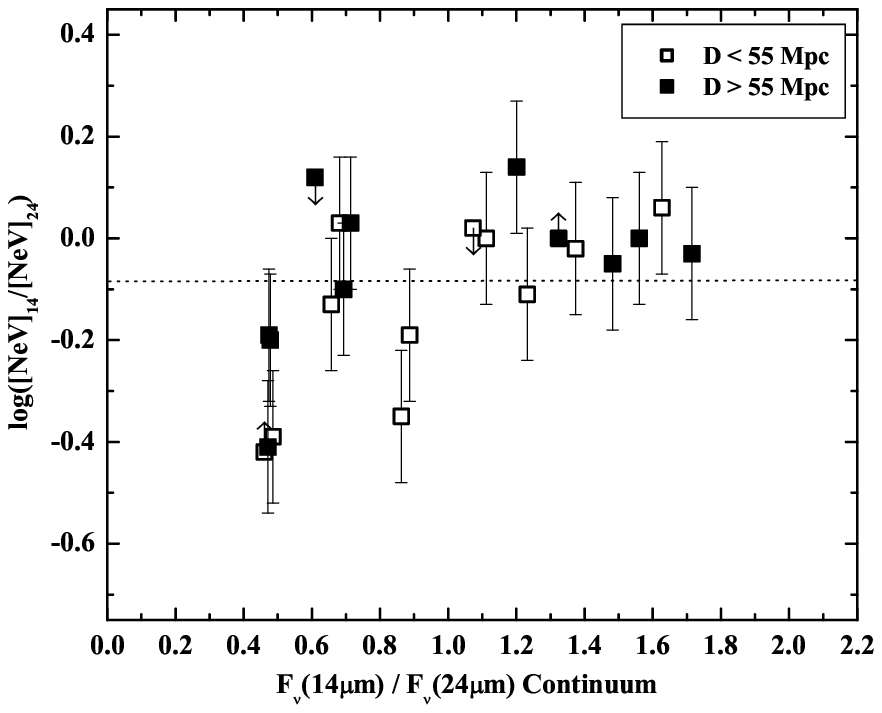}}\\
\end{center}
\caption[]{The [NeV] line ratio vs. the F$_{\nu}$(14$\mu$m)/F$_{\nu}$(24$\mu$m) continuum ratio for our sample. In both plots, the error bars mark the calibration uncertainties on the line ratio. There is a correlation between the line and continuum ratios which suggests that extinction affects the observed line flux ratios. 6a) The majority of galaxies with ratios below the LDL are Type 2 objects, implying that the extinction toward the [NeV]-emitting region may be due to the torus. 6b) The correlation shown here is not an artifact of aperture vatiations between the SH and LH slits.  The correlation holds when only the most distant galaxies are considered.}
\end{figure*}

The Spearman rank correlation coefficient for Figure 6 is 0.60 (with a probability of chance correlation of 0.008), indicating a significant correlation between the [NeV] line flux ratios and the mid-IR continuum ratio.  We note that some AGNs are known to contain prominent silicate emission features (Hao et al. 2005, Sturm et al. 2006) which have not been disentangled from the underlying continuum in this study.  Because of this, the 14$\mu$m or 24$\mu$m flux may be overestimated in some cases making intrinsic value of the continuum at 14$\mu$m and 24$\mu$m somewhat uncertain.  However only one galaxy plotted in Figure 6 is currently known to contain such features (PG1211+143, Hao et al. 2005).  Variations in n$_e$ and the underlying continuum shape will also add scatter to the correlation, as will differences in extinction to the line- and continuum-producing regions.  We should note that the correlation seen in Figure 6 is not an artifact of aperture variations between the LH and SH slit.  The correlation holds when only the most distant galaxies (closed symbols in Figure 6b) are considered.

Independent of this correlation, {\it our most important finding is that the [NeV] line flux ratio is significantly lower for Type 2 AGNs than it is for Type 1 AGNs}.  Figure 7 shows the relative [NeV] flux ratios for the Type 1 and Type 2 objects in our sample.  The mean ratios are 0.97 and 0.72 for the eight Type 1 and ten Type 2 AGNs, respectively, with uncertainties in the mean of about 0.08 for each.  Interestingly, although the sample size is limited, precluding us from drawing firm statistically significant conclusions, there is a similar suggestive trend seen in the sample of AGNs observed by Sturm et al. (2002) with {\it ISO}-SWS.  That is, in their work, the two galaxies with the lowest [NeV] flux ratios are NGC 1365 and NGC 7582, both Type 2 AGNs.  The galaxy with the highest ratio in their work is TOL 0109-383 , a Type 1 AGN.

\begin{figure}[htbp]
{\includegraphics[width=9cm]{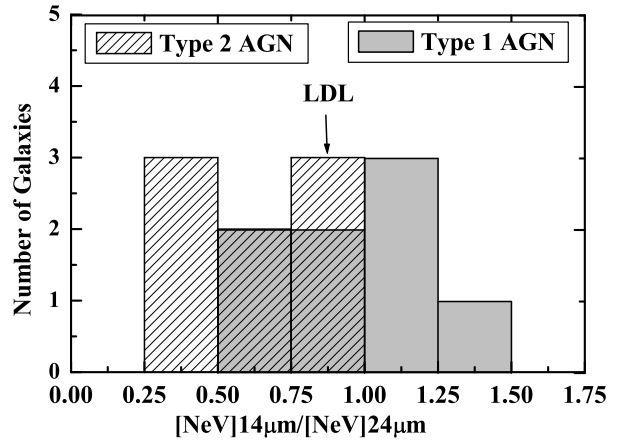}}\\
\caption[]{Histogram of the [NeV] 14$\mu$m/24$\mu$m line flux ratio as a function of AGN type.  The [NeV] line ratios for Type 2 AGNs are consistently lower than those from Type 1 AGNs.}
\end{figure}
If indeed the torus obscures the IR [NeV] emission in Type 2 objects, one would expect the optical/UV [NeV] emission in these objects to be obscured as well.   We searched the literature for optical/UV detections of [NeV] $\lambda$3426 for all of the galaxies in our sample and found five galaxies with observations at this wavelength.  Four of these galaxies (Mrk 463, Mrk 3, NGC 1566, and NGC 4151) were detected at [NeV] $\lambda$3426; the other (NGC 3031) was not detected (see, Kuraszkiewicz et al. 2002, 2004 and Forster et al. 2001 for optical/UV fluxes).  Of the four galaxies with optical/UV [NeV] detections, two are Type 1 galaxies (NGC 1566 \& NGC 4151) and, surprisingly, two are Type 2 galaxies (Mrk 3 and Mrk 463).  If the Type 2 galaxies Mrk 3 and Mrk 463 had [NeV] emitting regions interior to the torus, then the optical/UV lines in these objects should not be detected due to severe obscuration.  We note that Mrk 3 and Mrk 463 have some of the highest X-ray luminosities (both $\sim$ 10$^{43}$ erg s$^{-1}$ ) in the sample and mid-IR [NeV] ratios that are comparable to similarly luminous Type 1 objects--consistent with little or no obscuration in the mid-IR for these Type 2 galaxies.  This finding may imply that, in the most powerful AGNs, the [NeV] emitting region is pushed beyond the torus because the radiation field is so intense, while lines with higher ionization potentials than [NeV]  (such as [NeVI], [FeX] etc.) are still concealed by the torus.  More data, both from the mid-IR and from the optical/UV, are needed to further test this hypothesis.

\subsection{The [NeV]-emitting region is obscured by the host galaxy or dust in the NLR.}  

While the correlation in Figures 6a and 6b are very promising explanations for the observed [NeV] line ratios, it is not completely clear why some Type 2 galaxies appear obscured and others may not.  Perhaps the [NeV] emission is attenuated by dust in the NLR itself or elsewhere in the host galaxy.  Indeed, it is well-known that dust does exist in the NLR (e.g., Radomski et al. 2003, Tran et al. 2000) and that it can be extended and patchy (e.g Alloin et al. 2000; Galliano et al. 2005; Mason et al. 2006).  In addition, dust in the host galaxy could be responsible for the extinction seen here.  For completeness, we have conducted a detailed archival analysis of all of the galaxies in our sample with [NeV] ratios close to or below the LDL in order to see if there is additional evidence for high extinction either in the host galaxy or within the NLR.  We find that the majority of galaxies with low densities do indeed have well-known dust lanes, large X-ray inferred column densities, or other properties indicative of extinction. 

{\bf Cen A:} This nearby (D = 3.4Mpc) early type (S0) galaxy at one time devoured a smaller gas-rich spiral galaxy (Israel 1998, Quillen 2006).  There is clear evidence for substantial obscuration toward the nucleus of Cen A.  For example, the central region is veiled by a well known dense dust lane thought to be a warped thin disk (Ebneter \& Balick 1983, Bland et al. 1986, 1987, Nicholson et al. 1992, Sparke 1996, Israel 1998, Quillen et al. 2006).  Schreier et al. (1996) find V-band extinction averaging ~4-5 mag and infrared observations by Alonso \& Minniti yield A$_V$ values exceeding 30 mag in some regions. Thus, it is plausible that there are regions toward the nucleus of Cen A that are obscured even at infrared wavelengths.

{\bf NGC 1566:} The optical nuclear spectrum of this nearby galaxy is known to vary dramatically over a period of months, changing its optical classification from a Type 2 object to a Type 1 object and back again (Pastoriza \& Gerola 1970, de Vaucouleurs 1973, Penfold 1979, Alloin et al. 1985).   The narrow optical lines in this object also show prominent blue wings and the radio properties of this galaxy are more consistent with a Type 2 object than a Type 1 object (Alloin et al. 1985).  HST continuum imagery reveals spiral dust lanes within 1'' of the nucleus (Griffiths et al. 1997) which might be responsible for the Type 1/Type 2 variability.  Baribaud et al. (1992) find hot dust which lies just outside the broad line region in this galaxy and a large covering factor that might explain the steep continuum of the AGN.    Ehle et al. (1996) find N$_H$ $\sim$ 2.5 $\times$ 10$^{20}$ cm$^{-2}$ from ROSAT X-ray observations of this galaxy.  

{\bf NGC 2623:} This galaxy's tidal tails are evidence of a merger event, however infrared observations reveal a single symmetric nucleus, implying that the merging galaxies have coalesced.  Multi-color, near infrared observations reveal strong concentrations of obscuring material in the central 500 pc.(Joy \& Harvey 1987; Lipari et al. 2004).  Lipari et al. (2004) also find an optically obscured nucleus with V-band extinction $\geq$ 5 mag.

{\bf IRAS 23128-5919:}  This galaxy is also in the late stages of a merger.  The nuclei of the two galaxies are 4kpc apart and have not yet coalesced.  The northern nucleus is a starburst.  The southern nucleus is a known AGN, though its optical classification, Seyfert or LINER, is unclear (Duc, Mirabel, \& Maza 1997; Charmandaris et al. 2002; Satyapal et al. 2004).  IRAS 23128-5919 is an ultraluminous infrared galaxy (ULIRG), clearly consistent with the presence of substantial dust towards the nucleus. Optical spectroscopy of the southern nucleus indicates very large (1500 km s$^{-1}$) blue asymmetries in the H$\beta$ and [OIII] lines.  This blue wing could be a signature of extinction toward the far side of an expanding region, where the red wing is preferentially obscured. (Johansson \& Bergvall 1988).

{\bf Mrk 273:} This galaxy is also a ULIRG, so significant dust obscuration toward the nucleus is expected.  Near-IR imaging and high resolution radio observations show evidence for a double nucleus in this galaxy separated by less than 1 kpc (Ulvestad \& Wilson 1984; Mazzarella et al. 1991; Majewski et al. 1993).  However, high resolution Chandra observations reveal only the northern of the two nuclei, suggesting that this galaxy is hosting only one AGN and that perhaps the other "nucleus" is in fact a portion of the southern radio jet.  The soft X-ray emission from the northern nucleus is obscured by column densities of at least 10$^{23}$ cm$^{-2}$ (Xia et al. 2002).  Although the X-ray-emitting regions are physically distinct from the NLR and some of the obscuration at X-ray wavelengths likely arises in dust-free gas within the sublimation radius, the high column density derived may be consistent with high extinction toward the central regions of this galaxy.  Though Xia et al (2002) find that the X-ray morphology of the AGN in Mrk 273 is consistent with a Seyfert, Colina et al. (1999) find that it has a LINER optical spectrum, thus implying that some LINER galaxies are in fact heavily absorbed powerful AGN.  The soft diffuse X-ray halo in combination with the radio morphology found by Carilli \& Taylor (2000) may suggest a circumnuclear starburst surrounding the northern AGN nucleus, again consistent with substantial obscuration toward the AGN.  

{\bf NGC 3627:}  This nearby galaxy (D $\sim$ 10Mpc) is thought to have had tidal interactions with NGC 3628, a neighboring galaxy in the Leo Triplet, some 8 $\times$ 10$^8$ years ago which caused an intense burst of star formation in the nuclear regions around the same time (Rots 1978, Zhang et al. 1993, Afanasiev \& Sil'chenko 2005).  Zhang et al. (1993) also discovered an extremely dense molecular bar (mass $\geq$ 4 $\times$ 10$^8$ M$_{\odot}$) and Chemin et al. (2003) uncovered a warped disk using H$\alpha$ observations,  both evidence of the tidal interaction.  In their spectral fitting to the BeppoSAX observation of NGC 3627, Georgantopoulos et al. (2002) find intrinsic column densities of $\sim$ 1.5 $\times$ 10$^{22}$ cm$^{-2}$ which, like Mrk 273, may suggest substantial extinction to other regions near the nucleus.

{\bf NGC 7469:} This is a well-known, extensively-studied galaxy with strong, active star formation surrounding a Seyfert 1 nucleus.  Meixner et al. (1990) find dense molecular gas (2 $\times$ 10$^{10}$ M$_{\odot}$), two orders of magnitude above the Galactic value, within the central 2.5kpc of the nucleus.  3.3$\mu$m imaging of the galaxy reveals that 80\% of the PAH emission comes from an annulus $\sim$ 1''-3''in radius around the central nucleus, indicating that there is an elongated region of material that shelters the PAH from the harsh radiation field of the AGN (Cutri et al. 1984, Mazzarella et al. 1994).  [OIII] line asymmetries may corroborate the presence of a dense obscuring medium, revealing a blue wing resulting when the redshifted gas is obscured by the star forming ring (Wilson et al. 1986).  In addition, Genzel et al. (1995) find variation in the NIR emission attributable to extinction and estimate the extinction from the CO observations of Meixner et al. (1990) to be A$_V$ $\sim$ 10 mag. 

{\bf NGC 1365:} This nearby (D = 18.6 Mpc) AGN is known to be circumscribed by embedded young star clusters.  The galaxy also contains a prominent bar with a dust lane that penetrates the nuclear region (Phillips et al. 1983, Lindblad et al. 1996 \& 1999, Galliano et al. 2005).  Like NGC 7469, NGC 1365 shows a peak at 3.5$\mu$m implying PAH emission in spite of the harsh AGN radiation field (Galliano et al. 2005).  The large H$\alpha$/H$\beta$ ratio found by Alloin et al. (1981) implies substantial extinction toward the emission line regions, ranging from 3-4 mag.   Observations with ASCA and ROSAT imply high intrinsic column densities toward the X-ray emitting regions, suggesting possibly high obscuration towards other regions near the nucleus (Iyomoto et al. 1997, Komossa \& Schulz (1998), see also Schulz et al. 1999).  Komossa \& Schulz show that the ratio of H$\alpha$ to both the mid-IR and X-ray radiation is substantially different in NGC 1365 compared with typical Seyfert 1 galaxies, possibly suggesting inhomogenous obscuration (Schultz et al. 1999).  In an XMM X-ray study of NGC 1365, Risaliti et al. (2005) also find a heavily absorbed Seyfert nucleus.  The blueshifted X-ray spectral lines imply high column densities of 10$^{23}$ cm$^{-2}$ or more.  

\begin{figure}[htbp]
{\includegraphics[width=8cm]{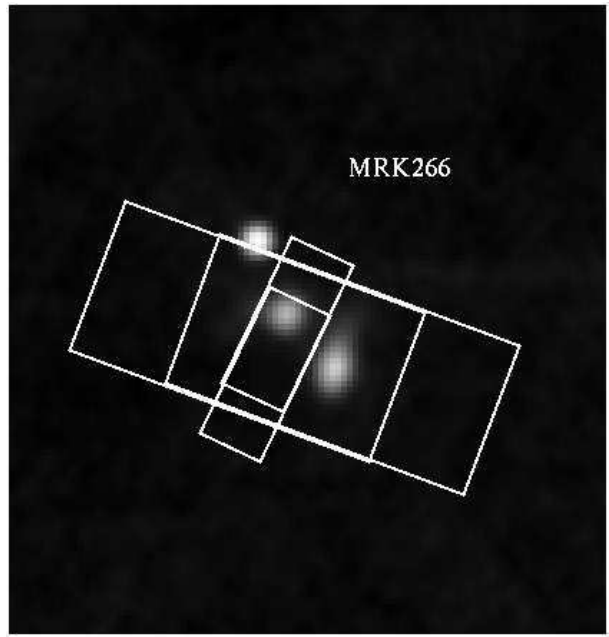}}\\
\caption[]{20 cm image of Mrk 266 taken from NED (http://nedwww.ipac.caltech.edu/).  As can be seen here, the SH slit (from which the 14$\mu$m line is extracted) overlaps with a third radio source, while the LH slit (from which the 24$\mu$m line is extracted) encompasses the southwestern nucleus and part of the northeastern nucleus.}
\end{figure}

{\bf Mrk 266 (NGC 5256):}  This luminous infrared galaxy is the only galaxy for which aperture effects most likely account for the low 14$\mu$m/24$\mu$m ratio.  Mrk 266 contains a very complicated structure which includes at least two bright nuclei, a Seyfert and a LINER, that are 10'' apart--a signature of a merger in progress.  The morphology of the northeast LINER nucleus is extremely controversial (Wang et al. 1997; Kollatschny \& Kowatsch 1998; Satyapal 2004, 2005; Ishigaki et al. 2000; Davies, Ward, \& Sugai 2000).  Mazzarella et al. (1988) find three non-thermal radio structures, two that coincide with the nuclei and one between the two nuclei.  Mazzarella et al. (1988) suggest that the two nuclear structures are associated with classical AGN and are in the stage of a violent interaction in which the center of gravity of the collision produces a massive burst of star formation with supernovae or shocks which are responsible for the third nonthermal radio source.  As can be seen in Figure 8, the SH slit, which provides the 14$\mu$m flux, overlaps with this third radio source, while the LH slit, responsible for the 24$\mu$m flux, encompasses the southwestern nucleus, the third radio source, and part of the northeastern nucleus.  In this case the two lines observed originate in physically distinct regions that do not each encompass all potential sources of [NeV] emission, resulting in an unphysical 14$\mu$m/24$\mu$m ratio.  This is not to say that Mrk 266 does not suffer from extinction at all.  Indeed the possible presence of a circumnuclear starburst implies that there may be substantial extinction (Ishigaki et al. 2000; Davies, Ward, \& Sugai 2000).  We have verified that this is the only distant galaxy in our sample with a complicated nuclear structure that will result in aperture effects.

\subsection{Can the [NeV] line flux ratio be used as a density diagnostic?}

Our analysis reveals that extinction towards parts of the NLR in some objects is significant and cannot be ignored at mid-IR wavelengths.  In fact, it is quite possible that extinction affects the [NeV] line flux ratios of those galaxies with ratios above the low density limit (LDL) and the amount of extinction in all cases is highly uncertain.  In addition to extinction, the temperature of the [NeV] emitting gas is unknown.  If the [NeV] emission originates within the walls of the obscuring central torus, which may be the source of extinction in many of our galaxies, we might expect the temperature of the gas to reach 10$^6$ K (Ferland et al. 2002).  If, on the other hand, the [NeV] emission comes from further out in the NLR and is instead attenuated by the intervening material, we might expect the temperature of the gas to be closer to 10$^4$ K.  As shown in Figure 1, the electron densities inferred from the [NeV] line flux ratios are sensitive to temperature when such large temperature variations are considered. Based on the calculations shown in Figure 1, the low ratios could indicate that the densities in the [NeV] line emitting gas are typically $\leq$ 3000 cm$^{-3}$ for T = 10$^{4} \K$.  However, if the [NeV] gas is characterized by temperatures as high as T = 10$^{5} \K$ to 10$^{6} \K$, densities as high as 10$^{5}$ cm$^{-3}$ would be consistent with our measurements.   We note that the [NeV] line flux ratios for the galaxies in our sample (especially the Type 1 AGNs) all cluster around a ratio of $\approx$ 1.0.  Two separate conclusions may be drawn from this finding: 1) That the temperatures of the gas are low ($\sim$ 10$^4 \K$) and that the electron density is relatively constant over many orders of magnitude in X-ray Luminosity and Eddington Ratio for these AGNs, or 2) That the temperature of the gas is high (10$^{5} \K$ to 10$^{6} \K$) and that the AGNs here sample a wide range of electron densities (from 10$^{2}$ cm$^{-3}$ to 10$^{5}$ cm$^{-3}$).  Since gas temperature, electron density, mid-IR continuum, and extinction are all unknown for these objects, the electron density cannot be determined here.

\section{The SIII Line Flux Ratios}

In Figure 9 we plot the $18\micron/33\micron$ line ratio as a function of electron density $n_e$.  As with [NeV], we only consider the five levels of the ground configuration when computing the line ratio and we plot the relationship for gas temperatures of $T = 10^4 \K$ and $10^5$K.  We adopt collision strengths from Tayal \& Gupta (1999) and radiative transition probabilities from Mendoza \& Zeippen (1982).  

\begin{figure}[htbp]
{\includegraphics[width=8cm]{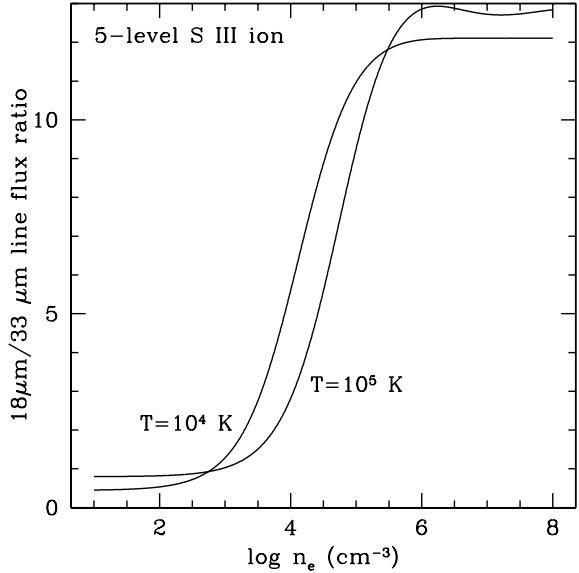}}\\
\caption[]{$18\micron/33\micron$ line flux ratio in S III versus electron density $n_e$, for gas temperatures $T=10^4 \K$ and $10^5 \K$.}
\end{figure}

In Table 3, we list the observed [SIII] line flux ratios for the galaxies in our sample.  As with the [NeV] ratios, the [SIII] ratios in many galaxies listed in Table 3 are well below the theoretically allowed value of 0.45 for a gas temperature of $T = 10^4 \K$ (13/33 detections). Again we explore the observational effects and the theoretical uncertainties that could artificially lower these ratios.

{\bf Aperture Effects:} The ionization potential of [SIII] is $\sim$ 35 eV and therefore the [SIII] emission may arise from gas ionized by either the AGN or young stars.  In Table 3 we list, in addition to our {\it Spitzer} [SIII] fluxes, all available [SIII] fluxes from ISO.  Unlike [NeV], the [SIII] fluxes from ISO are significantly larger than the {\it Spitzer} fluxes for most galaxies.   In Figure 10 we plot the {\it ISO} to {\it Spitzer}  flux ratios for the 18$\mu$m and 33$\mu$m the [SIII] lines.  As can be seen here, the [SIII] emission extends beyond the {\it Spitzer} slit for many galaxies (6 out of 9 for [SIII] 18$\mu$m and 11 out of 13 for [SIII] 33$\mu$m).  Similarly, when we compare the [SIII] flux arising from a single slit centered on the nucleus to the flux arising from a more extended region obtained using mapping observations (Dale et al.  2006), we find that in most cases the fluxes from the extended region are much larger than the nuclear single-slit fluxes.  Galaxies with fluxes from Dale et al. (2006) are not included in Figure 10 since the extraction aperture for these galaxies is comparable to the 18$\mu$m {\it ISO} slit. We point out that the value for this ratio is dependent on the orientation of the {\it Spitzer} slit relative to the {\it ISO} slit and on the distance of each object.  We also note that IRAS20551 and IRAS23128 are point sources with {\it Spitzer} 18$\mu$m fluxes greater than the {\it ISO} fluxes from Genzel et al. (1998), however they fall within the Genzel et al. (1998) quoted errors of 30\% and the Spitzer calibration error of 15\%.  Figure 10 suggests that the [SIII] emission may be produced in the extended, circumnuclear star forming regions associated with many AGNs and that {\it aperture effects need to be considered in our analysis of the [SIII] ratio for nearby objects.}

\begin{figure}[htbp]
{\includegraphics[width=9cm]{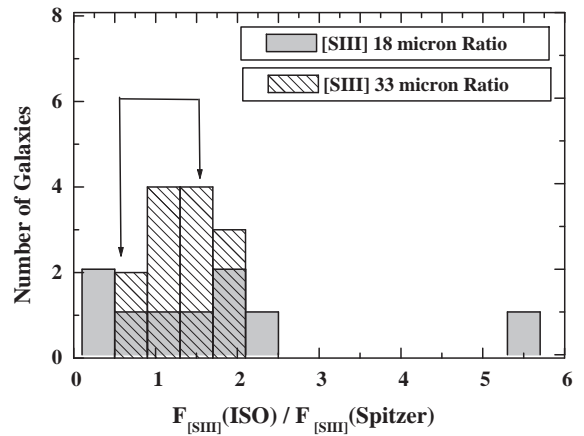}}\\
\caption[]{The ratios of the [SIII] flux from {\it ISO} and {\it Spitzer} for the 18$\mu$m and 33$\mu$m lines.  The range indicated with arrows is that corresponding to the absolute flux calibration for {\it ISO} (20\%) and {\it Spitzer} (15\%).  The [SIII] emission is indeed extended beyond the {\it Spitzer} slit for many galaxies, suggesting that the [SIII] emission may be produced in star forming regions. We note that IRAS20551 and IRAS23128 are point sources with {\it Spitzer} 18$\mu$m fluxes greater than the {\it ISO} fluxes from Genzel et al. 1998, however they fall within the Genzel et al. (1998) quoted errors of 30\% and the Spitzer calibration error of 15\%.  Galaxies with fluxes from Dale et al. 2006 are not included in this plot since the extraction aperture for these galaxies is comparable to the 18$\mu$m {\it ISO} slit.}
\end{figure}

  The contribution from star formation to the [SIII] lines can be estimated using the strength of the PAH emission, one of the most widely used indicators of the star formation activity in galaxies (e.g. Luhman et al. 2003; Genzel et al. 1998; Roche et al. 1991; Rigopoulou et al. 1999, Clavel et al. 2000; Peeters, Spoon, \& Tielens 2004).  We examined the [SIII] 18.71 $\mu$m/PAH 6.2 $\mu$m and [SIII] 33.48 $\mu$m/PAH 6.2 $\mu$m line flux ratios in 7 starburst galaxies observed by {\it Spitzer} and found them to be comparable to the analogous ratios in our entire sample of AGNs as shown in Figure 11.  This suggests that the bulk of the [SIII] emission originates in gas ionized by young stars.  We note that the apertures of the SH and LH IRS modules are smaller than that of the SL2 module, which may artificially raise the line ratios plotted in Figure 11 for nearby galaxies compared with the more distant ones.  However, the fact that the line ratios plotted in Figure 11 span a very narrow range suggests that the [SIII] line emission has a similar origin in starbursts and in AGNs. Thus, we assume that the bulk of the [SIII] emission originates in gas ionized by young stars and that the electron densities derived using these lines taken from slits of the same size (such those galaxies coming from Dale et al. 2006 mapping observations) or from the most distant galaxies are representative of the gas density in star forming regions.  

\begin{figure*}[htbp]
\begin{center}
{\includegraphics[width=9cm]{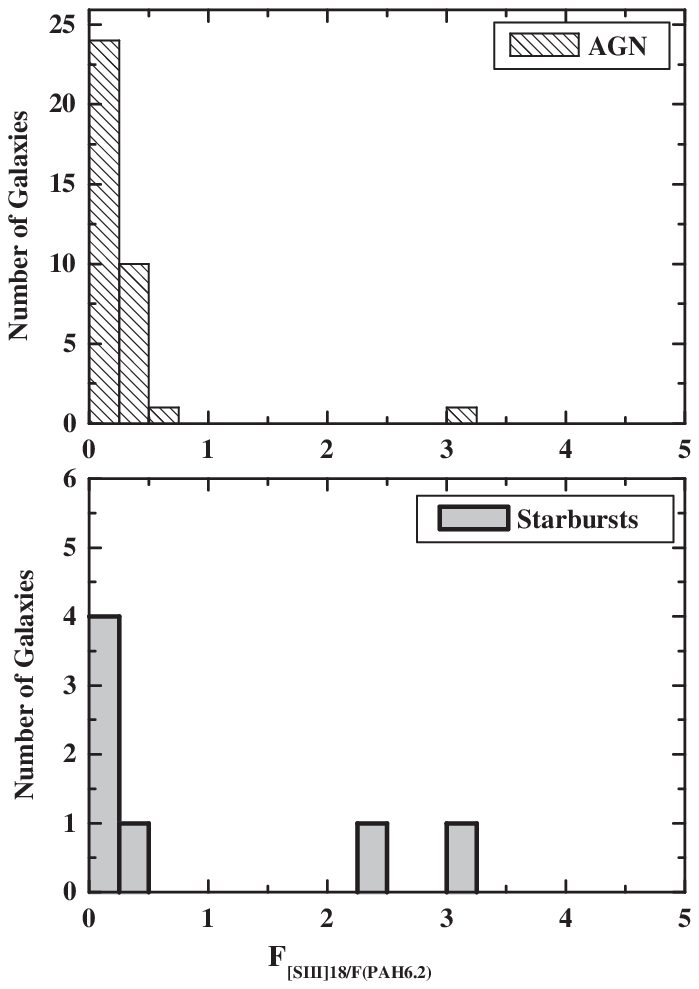}}{\includegraphics[width=9cm]{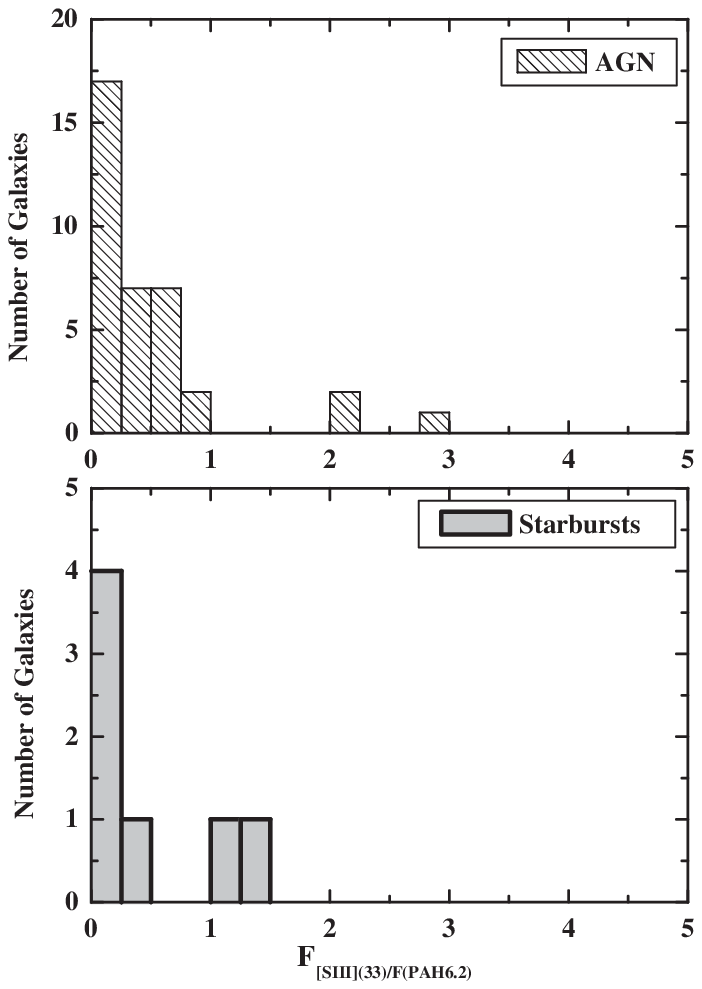}}\\\end{center}
\caption[]{Distribution of the [SIII]33$\mu$m/PAH 6.2$\mu$m and the [SIII]18$\mu$m/PAH 6.2$\mu$m line flux ratios for our sample of AGNs and a small sample of starburst galaxies observed by {\it Spitzer}.  It is apparent that the line ratios of the AGNs are comparable to the corresponding ratios in starbursts, suggesting that the bulk of the [SIII] emission originates in star forming regions and not the NLRs in our sample of AGNs. }
\end{figure*}

{\bf Extinction:} We have shown that aperture effects are the likely explanation for why many of the [SIII] ratios for the galaxies in our sample fall below the LDL.  However, there are three galaxies in the sample with ratios below the LDL that are distant enough (D$>$55 Mpc, corresponding to projected distances greater than 1.2 by 3 kpc and 3 by 6 kpc for the SH and LH slits, respectively) that aperture effects may not be as important (NGC 2623 \& Mrk 273, Mrk 266 has been excluded since it is known to be affected by aperture variations See Section 5.2).  Extinction may be the explanation for the low ratios in these galaxies.  However, even though the SH and LH slits likely cover the entirety of the NLR at these distances, we note that these three galaxies contain well-known, large circumnuclear starbursts (See Section 5.2 for the individual galaxy summaries) which may produce extremely extended [SIII] emission. It is therefore still possible that the line ratios in these galaxies are artificially lowered by aperture variations between the SH and LH slits.  However, in addition to these three distant galaxies, NGC 4725 from Dale et al. (2006) has a [SIII] ratio below the low density limit.  The low [SIII] ratio ($<$0.45) in this case cannot be attributed to aperture variations since the extraction region is the same for both the 18 and 33$\mu$m lines.  Thus, for completeness, the extinction derived using the extinction curves given in Section 4 from the observed [SIII] line ratio for these four sources are given in Table 3.  The Draine (1989) and Lutz et al. (1996) extinction curve calculations yield extinction values that range from $\sim$ 12 to 25 mag.  The Chiar and Tielens (2006) extinction curve for the Galactic Center may also be used since, unlike [NeV], the extinction at the longer wavelength line (33$\mu$m) is greater than that at the shorter wavelength line (18$\mu$m).  The values derived from this method are quite similar, ranging from $\sim$ 10 to 22 mag.  The Chiar and Tielens (2006) extinction curve from the local ISM cannot be used here since it only extends to 27.0$\mu$m.  

{\bf Computed Quantities:} As with NeV, there may be uncertainties in the computed SIII infrared collisional rate coefficients.  However, there is generally less controversy surrounding the [SIII] coefficients and these values are widely accepted.  

Our analysis suggests that aperture effects severely influence the [SIII] line flux ratios in most cases and that the observed flux is likely dominated by star forming regions.  Figure 12, a plot of the [SIII] line ratio as a function of distance, illustrates the influence of aperture effects on the [SIII] line ratio.  Most of the galaxies at distances $<$55 Mpc with [SIII] fluxes extracted from apertures of different sizes (i.e. NOT the Dale et al. (2006) galaxies) are below the LDL.  On the other hand, galaxies at larger distances and galaxies with fluxes from Dale et al. (2006) are generally above the LDL.  Thus, for the most distant galaxies in our sample and the galaxies with fluxes from Dale et al. (2006) where the aperture for the 18 and 33 $\mu$m lines are equal, aperture effects are not problematic, but extinction, as can be seen from Mrk 273, NGC2623, and NGC 4725 in Figure 12, needs to be considered.  As with the [NeV] line ratio, the [SIII] line ratio is NOT a tracer of the electron density in our sample.  In conclusion, the ambiguity of the intrinsic [SIII] line ratio is primarily the result of aperture variations.  However there is at least one case (NGC 4725) where aperture effects cannot explain the low ratio, implying that, in addition to aperture variations, extinction likely plays a role in lowering the [SIII] line flux ratios.

\begin{figure}[htbp]
{\includegraphics[width=9cm]{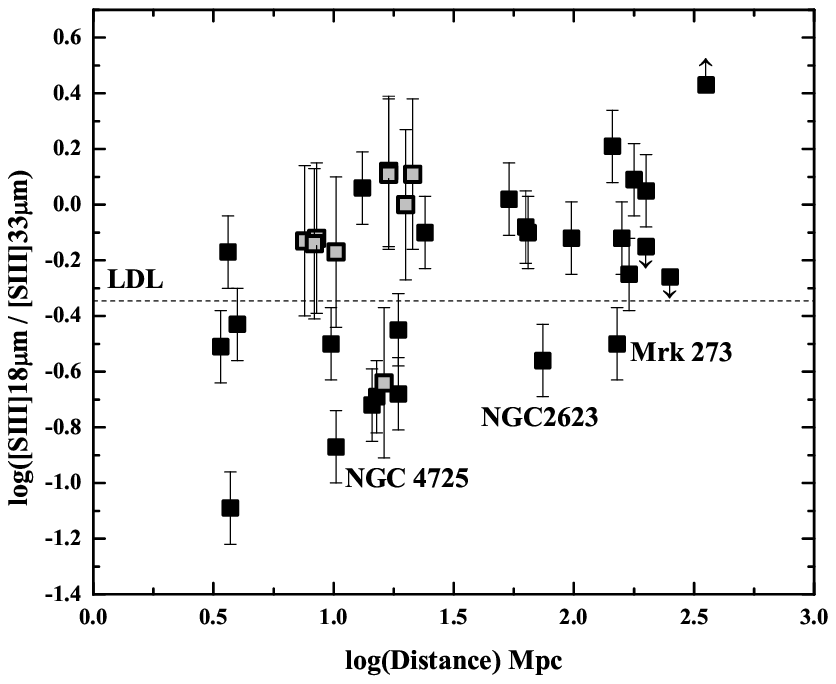}}\\
\caption[]{The [SIII] 18$\mu$m/33$\mu$m line ratio as a function of distance.  The error bars shown here mark the calibration uncertainties on the line ratio. Dale et al. (2006) quote 30\% calibration error which is shown here for those galaxies.  For the rest of the sample the calibration error is 15\% as per the Spitzer handbook.  Most of the galaxies at distances $<$55 Mpc with [SIII] fluxes extracted from apertures of different sizes (i.e. not the Dale et al. (2006) galaxies) are below the LDL.  However, for the most distant galaxies in our sample and the galaxies with fluxes from Dale et al. (2006) where the aperture for the 18 and 33 $\mu$m lines are equal, aperture effects are not problematic, but extinction needs to be considered (see  Mrk 273, NGC2623, and NGC 4725 above). }
\end{figure}

\section{Summary}

We report in this paper the [NeV] 14$\mu$m/24$\mu$m and [SIII]18$\mu$m/33$\mu$m line flux ratios, traditionally used to measure electron densities in ionized gas, in an archival sample of 41 AGNs observed by the {\it Spitzer Space Telescope}. 

\begin{enumerate}

\item We find that the [NeV] 14$\mu$m/24$\mu$m line flux ratios are low: approximately 70\% of those measured are consistent with the low density limit to within the calibration uncertainties of the IRS.	

\item We find that Type 2 AGNs have lower [NeV] 14$\mu$m/24$\mu$m line flux ratios than Type 1 AGNs.   The mean ratios are 0.97 and 0.72 for the eight Type 1 and ten Type 2 AGNs, respectively, with uncertainties in the mean of about 0.08 for each.

\item  For several galaxies, the observed [NeV] line ratios are below the theoretical low density limit.  All of these galaxies are Type 2 AGNs.

\item We discuss the physical mechanisms that may play a role in lowering the line ratios:  differential mid-IR extinction, low density, high temperature, and high mid-IR radiation density.

\item We argue that the [NeV]-emitting region likely originates interior to the torus in many of these AGNs and that differential infrared extinction due to dust in the obscuring torus may be responsible for the ratios below the low density limit.  We suggest that the ratio may be a tracer of the torus inclination angle to our line of sight.

\item Our results imply that the extinction curve in these galaxies must be characterized by higher extinction at 14$\mu$m than at 24$\mu$m, contrary to recent studies of the extinction curve toward the Galactic Center.

\item A comparison between the [NeV] line fluxes obtained with {\it Spitzer} and {\it ISO} reveals that there are systematic discrepancies in calibration between the two instruments.  However, our results are independent of which instrument is used; [NeV] line flux ratios are consistently lower in Type 2 AGNs than in Type 1 and [NeV] line flux ratios below the LDL are observed with both ISO and Spitzer.

\item Our work provides strong motivation for investigating the mid-IR spectra of a larger sample of galaxies with {\it Spitzer} in order to test our conclusions.       

\item Finally, an analysis of the [SIII] emission reveals that it is extended in many or all of the galaxies and likely originates in star forming gas and NOT the NLR.  Since there is a variation in the apertures between the SH and LH modules of the IRS, we cannot use the [SIII] line flux ratios to derive densities for the majority of galaxies in our sample.

\end{enumerate}
We are extremely thankful for all of the invaluable data analysis assistance from Dan Watson and Joel Green, without which this work would not have been possible.  We are also very grateful to Davide Donato, Eli Dwek, Frederic Galliano, Paul Martini, Kartik Sheth, Eckhard Sturm, Peter van Hoof, and Dan Watson for their enlightening and thoughtful comments/expertise that significantly improved this paper.   Carissa Khanna was also very helpful in providing assistance in the preliminary data analysis.  We are also grateful for the helpful and constructive comments from the referee.  This research has made use of the NASA/IPAC Extragalactic Database (NED) which is operated by the Jet Propulsion Laboratory, California Institute of Technology, under contract with the National Aeronautics and Space Administration.  SS gratefully acknowledges financial support from NASA grant NAG5-11432 and NAG03-4134X.  JCW gratefully acknowledges support from Spitzer Space Telescope Theoretical Research Program.  JCW is a Cottrell Scholar of Research Corporation.  Research in infrared astronomy at NRL is supported by 6.1 base funding.  RPD gratefully acknowledges financial support from the NASA Graduate Student Research Program.



\end{document}